\documentclass{article}

\usepackage{arxiv}

\usepackage[utf8]{inputenc} % allow utf-8 input
\usepackage[T1]{fontenc}    % use 8-bit T1 fonts
\usepackage{hyperref}       % hyperlinks
\usepackage{url}            % simple URL typesetting
\usepackage{booktabs}       % professional-quality tables
\usepackage{amsfonts}       % blackboard math symbols
\usepackage{nicefrac}       % compact symbols for 1/2, etc.
\usepackage{microtype}      % microtypography
\usepackage{lipsum}
\usepackage{amsmath}
\usepackage{pgf,tikz,pgfplots}

\usepackage{color}
\usepackage[utf8]{inputenc}
\usepackage[T1]{fontenc}
\usepackage[caption=false]{subfig}
\usepackage{notoccite}% for arranging citations
\usepackage{longtable}
\usepackage{pgf,tikz,pgfplots}
\usepackage{mathrsfs}
\usetikzlibrary{arrows}
\usepackage{enumitem}
\usepackage{graphicx}% Include figure files
\usepackage{dcolumn}% Align table columns on decimal point
\usepackage{bm}% bold math

\usepackage{url}

\usepackage{color}

\usepackage[utf8]{inputenc}
\usepackage[T1]{fontenc}
\pgfplotsset{compat=1.14}

\newcommand{\mysize}{0.45}
\newcounter{undefinedreferences}
\newcommand*\midpoint[1]{\overline{#1}}
\title{Refined betatron tune measurements by mixing beam position data}% Force line breaks with \\

\author{
 Panos~Zisopoulos\thanks{Use footnote for providing further
    information about author (webpage, alternative
    address)---\emph{not} for acknowledging funding agencies.} \\
  CERN\\
  Geneva, CH-1211 \\
  Switzerland \\
  \texttt{panagiotis.zisopoulos@cern.ch} \\
   \And
 Yannis~Papaphilippou \\
  CERN\\
  Geneva, CH-1211 \\
  Switzerland \\
   \AND
   Jacques Laskar \\
   IMCCE, Observatoire de Paris\\
   77 Avenue Denfert-Rochereau\\
   75014 Paris \\
}
\begin{document}
\maketitle
% \preprint{APS/123-QED}

% \thanks{A footnote to the article title}%

% \author{P. ~Zisopoulos}
% \email{panagiotis.zisopoulos@cern.ch}
%  \altaffiliation[Also at ]{High Energy Physics Department, Uppsala University.}%Lines break automatically or can be forced with \\
% \author{Y.~Papaphilippou}%
 % \email{jacques.laskar@institution.edu}

% \affiliation{%
% CERN, CH-1211, Geneva, Switzerland
%  % \textbackslash\textbackslash
% }%

% \collaboration{MUSO Collaboration}%\noaffiliation

% \author{J.~Laskar}
% % \homepage{http://www.Second.institution.edu/~Charlie.Author}
% \affiliation{
%  IMCCE, Observatoire de Paris\\77 Avenue Denfert-Rochereau\\75014 Paris% with \\
% }%

% \collaboration{CLEO Collaboration}%\noaffiliation

\date{\today}% It is always \today, today,
             %  but any date may be explicitly specified

\begin{abstract}
The measurement of the betatron tunes in a circular accelerator is of paramount importance due to their impact on beam dynamics. The resolution of the these measurements, when using turn by turn (TbT) data from beam position monitors (BPMs), is greatly limited by the available number of turns in the signal. Due to decoherence from finite chromaticity and/or amplitude detuning, the transverse betatron oscillations appear to be damped in the TbT signal. On the other hand, an adequate number of samples is needed, if precise and accurate tune measurements are desired. In this paper, a method is presented that allows for very precise tune measurements within a very small number of turns. The theoretical foundation of this method is presented with results from numerical and tracking simulations but also from experimental TbT data which are recorded at electron and proton circular accelerators.

% \begin{description}
% \item[Usage]
% Secondary publications and information retrieval purposes.
% \item[PACS numbers]
% May be entered using the \verb+\pacs{#1}+ command.
% \item[Structure]
% You may use the \texttt{description} environment to structure your abstract;
% use the optional argument of the \verb+\item+ command to give the category of each item.
% \end{description}

\end{abstract}

% \pacs{Valid PACS appear here}% PACS, the Physics and Astronomy
                             % Classification Scheme.
%Use showkeys class option if keyword
                              %display desired
% \maketitle

%\tableofcontents

% \maketitle

\section{\label{sec:level1}Introduction}

Betatron tune measurements~\cite{Serio:BFb0018282} are used as a reliable diagnostic of transverse beam dynamics. The measurement of the working point of a circular accelerator, is an essential procedure in order to optimize performance and reduce particle losses. In such an accelerator, each BPM records the betatron oscillations of the centroid in the transverse plane for many turns. In order to measure the tunes, the beam needs to perform coherent betatron oscillations after transverse excitation from its closed orbit. The TbT transverse oscillations of the centroid are recorded at each BPM, yielding a discrete signal which can be analyzed with algorithms that perform Fourier analysis.

The figure of merit in tune measurements, is the resolution in the frequency space, i.e. the smallest difference of adjacent harmonics that can be identified in the Discrete Fourier Transform (DFT) spectrum of a BPM signal. In the noise-free regime, the resolution is defined from the error $\epsilon(N) $ in the betatron tune estimation from $N$ turns,  which follows a power law of the form~\cite{Bartolini:1996gj}:
\begin{equation}
\label{law}
	\epsilon(N) =|Q(N)-Q_o| \propto \frac{1}{N^l}\,\,,
\end{equation}
where $Q(N)$ is the estimated tune within a number of turns $N$, $Q_o$ is the actual betatron tune and $l$ is an exponent which determines the speed of convergence of the tune measurement to the actual tune value.

As it is suggested from Eq.~\eqref{law}, a large number of turns is vital for precise betatron tune measurements. However, in experimental and simulated TbT data, the useful number of turns is greatly limited from decoherence due to finite chromaticity and/or amplitude detuning~\cite{Meller:1987ug, Lee:1991ii}. Due to this mechanism, the expectation value of the transverse beam position, as recorded at the BPMs, is strongly damped with respect to the number of turns. Consequently, a choice of a large number of turns is followed from a substantial reduction of the Signal-to-Noise Ratio (SNR) in the TbT signal. In addition, decoherence induces a significant dependence of the betatron tunes on the number of turns.
Apart from non-linearities, a large number of turns is not always available e.g. in the case of accelerator commissioning or due to particular beam dynamics measurements e.g. placing the working point of the accelerator close to a resonance that may lead to strong particle losses.

For the case of a simple Fast Fourier Transform (FFT) algorithm~\cite{Cooley:1965zz}, the exponent $l$ in Eq.~\eqref{law} is $l=1$, which is inadequate for precise determination of the betatron tunes in a circular accelerator.
To overcome such limitations, Refined Frequency Analysis (RFA) methods have been developed~\cite{Bartolini:1996gj,Asseo:193794,Gasior:2004md} which offer a substantially improved resolution in tune estimations, with respect to a plain FFT. The \emph{Numerical Analysis of Fundamental Frequencies} (NAFF)~\cite{Laskar:1992zz} algorithm is a RFA method, which has been widely used with success in beam dynamics measurements~\cite{Laskar:2003zz,Skokos:2004wn,Papaphilippou:2014jma,Zisopoulos:2014hva}. This algorithm offers an enhanced decrease of the error in Eq.~\eqref{law} after applying a Hann window~\cite{F.J.Harris:1978} of order $p$ on the data. The resolution has been analytically calculated from Laskar to have the asymptotic approximation for $N\to\infty$~\cite{LASKAR1993257}:
\begin{equation}
\label{law_naff}
\epsilon(N)  \propto \frac{1}{N^{2p+2}}\,\,,
\end{equation}
which offers a substantial improvement in the tune estimation, with respect to a simple FFT, for a given number of turns $N$. Nevertheless, a further enhancement of the resolution would be beneficial for the operation of modern circular accelerators and for analyzing TbT data from tracking simulations. The reason for this, is that the usage of a very small number of turns ($N\leq50$) results in a weaker influence of decoherence on the tune determination. In addition, the identification of transient effects, such as ripples from power converters or fast-pulsing magnets, could be possible.

In this paper, it is shown that by analyzing the trajectory of the beam from $M$ BPMs and for $N$ turns, while employing the NAFF algorithm, the resolution of the tune measurements is greatly improved, several orders of magnitude  compared to single BPM analysis. The reconstruction can be performed by vectorizing an $N \times M$ array that contains TbT data for  $N$ turns and from $M$ BPMs, in a BPM-by-BPM manner. This transformation results in the simultaneous increase of the number of samples ($MN$ from $N$) and of the sampling rate, since after the vectorization, the sampling rate is $M$ samples per turn, instead of the initial $1$ sample per turn.

 This technique is referred to as \emph{the mixed BPM method} and when it is employed, the TbT error in the tune estimation scales as:
 \begin{equation}
\label{law2}
\epsilon(N) \propto \frac{1}{M^{l-1}N^{l}}\,\,,
\end{equation}
where the gain in resolution is pronounced from the appearance of the $M^{l-1}$ factor, if compared to the single BPM analysis, Eq~\eqref{law}.

The resolution in tune estimations, while utilizing the NAFF algorithm and the mixed BPM method scales as:
\begin{equation}
\label{law_naff_mixed_0}
\epsilon(N) \propto \frac{1}{M^{2p+1}N^{2p+2}}\,\,,
\end{equation}
where the increase in the resolution of the tune measurement by a factor of $M^{2p+1}$ is evident.

The mixed BPM method consists of flattening the $MxN$ array in a BPM-by-BPM manner i.e., transforming it into a vector of $1\times MN$ dimension. However, the transformation introduces two systematic errors:
\begin{itemize}

\item An error in the sampling period of the mixed BPM signal due to the fact that the BPMs are not strictly equidistant, i.e. they are not distributed homogeneously around the ring.
\item An error due to the BPM-by-BPM modulation of the beta function.
\end{itemize}

Fortunately, the previous errors are \emph{periodic} in nature since they are repeated for every revolution of the beam. This characteristic allows for the aforementioned analysis of the trajectory of the beam in a BPM-by-BPM manner.

The mixed BPM method has been explored in the past at proton and electron rings \cite{Papaphilippou:2007zz,Zisopoulos:2015zkg} for precise tune measurements with a very small number of turns. Here, the theoretical foundations of this method is presented, along with results from tracking simulations and experimental measurements. The paper is organized as follows:

 In Section~\ref{sec:Methodology} the methodology of the mixed BPM scheme is presented, together with analytical expressions for the transformed  Fourier spectra and the improved betatron tune resolution. In Section~\ref{sec:NumApplication}, a numerical simulation is deployed in order to visualize qualitatively and quantitatively the theoretical results of the method. In Section~\ref{sec:Sims} tracking simulations are performed and the method is applied for tune measurements with the CERN Proton Synchrotron (PS) ideal lattice, under the influence of linear and non-linear dynamics. In Section~\ref{sec:Applications} some experimental results are shown from the PS and the CERN Proton Synchrotron Booster (PSB) and from the KARA electron light source, where the efficiency of the mixed BPM method is highlighted.

\section{\label{sec:Methodology}Methodology}

\subsection{Non-uniform periodic sampling\label{Method_sub1}}

\begin{figure}[htb!]
\centering
\definecolor{dtsfsf}{rgb}{0.8274509803921568,0.1843137254901961,0.1843137254901961}
\definecolor{rvwvcq}{rgb}{0.08235294117647059,0.396078431372549,0.7529411764705882}
\begin{tikzpicture}
\draw [line width=1pt] (0,0) circle (3cm);
\draw [line width=1pt,color=rvwvcq] (-2.0853550021597353,2.1566860028681436)-- (2.068318017144739,-2.1730302758945754);
\draw [line width=1pt,color=rvwvcq] (2.1197367239559193,2.1229027818326083)-- (-2.1213203435596424,-2.121320343559643);
\draw [line width=1pt,color=rvwvcq] (0,3)-- (0,-3);
\draw [line width=1pt,color=rvwvcq] (-3,0)-- (3,0);

\draw (-2.308181818181817,2.7809090909090605) node[anchor=north west] {\textbf{$\delta_8$}};
\draw (1.6372727272727283,2.9718181818181504) node[anchor=north west] {\textbf{$\delta_2$}};
\draw (3.0281818181818193,-0.23545454545454492) node[anchor=north west] {\textbf{$\delta_3$}};
\draw (2.4372727272727284,-1.7372727272727088) node[anchor=north west] {\textbf{$\delta_4$}};
\draw (-0.5809090909090896,-3) node[anchor=north west] {\textbf{$\delta_5$}};
\draw (-2.4536363636363625,-2.3463636363636148) node[anchor=north west] {\textbf{$\delta_6$}};
\draw (-3.49,-0.02636363636363693) node[anchor=north west] {\textbf{$\delta_7$}};

\draw [shift={(0,0)},line width=2pt,color=dtsfsf,fill=dtsfsf,fill opacity=0.5]  plot[domain=3.141592653589793:3.3714970817216976,variable=\t]({1*3*cos(\t r)+0*3*sin(\t r)},{0*3*cos(\t r)+1*3*sin(\t r)});
\draw [shift={(0,0)},line width=2pt,color=dtsfsf,fill=dtsfsf,fill opacity=0.5]  plot[domain=2.1254969270623936:2.3393808198975745,variable=\t]({1*3*cos(\t r)+0*3*sin(\t r)},{0*3*cos(\t r)+1*3*sin(\t r)});
\draw [shift={(0,0)},line width=2pt,color=dtsfsf,fill=dtsfsf,fill opacity=0.5]  plot[domain=0.7861444104647892:1.1820856352461153,variable=\t]({1*3*cos(\t r)+0*3*sin(\t r)},{0*3*cos(\t r)+1*3*sin(\t r)});
\draw [shift={(0,0)},line width=2pt,color=dtsfsf,fill=dtsfsf,fill opacity=0.5]  plot[domain=-0.26275944547639085:0,variable=\t]({1*3*cos(\t r)+0*3*sin(\t r)},{0*3*cos(\t r)+1*3*sin(\t r)});
\draw [shift={(0,0)},line width=2pt,color=dtsfsf,fill=dtsfsf,fill opacity=0.5]  plot[domain=5.473103721296365:5.695408431741635,variable=\t]({1*3*cos(\t r)+0*3*sin(\t r)},{0*3*cos(\t r)+1*3*sin(\t r)});
\draw [shift={(0,0)},line width=2pt,color=dtsfsf,fill=dtsfsf,fill opacity=0.5]  plot[domain=4.4364055801868405:4.71238898038469,variable=\t]({1*3*cos(\t r)+0*3*sin(\t r)},{0*3*cos(\t r)+1*3*sin(\t r)});
\draw [shift={(0,0)},line width=2pt,color=dtsfsf,fill=dtsfsf,fill opacity=0.5]  plot[domain=3.9269908169872414:4.113024185903882,variable=\t]({1*3*cos(\t r)+0*3*sin(\t r)},{0*3*cos(\t r)+1*3*sin(\t r)});

\draw [line width=1pt,color=dtsfsf] (0,0)-- (1.1369868191682282,2.7761954133377778);
\draw [line width=1pt,color=dtsfsf] (0,0)-- (-1.5800664498749424,2.550174506573932);
\draw [line width=1pt,color=dtsfsf] (-2.921064536389272,-0.6836534021336591)-- (0,0);
\draw [line width=1pt,color=dtsfsf] (0,0)-- (-1.692354305339511,-2.477082337185993);
\draw [line width=1pt,color=dtsfsf] (0,0)-- (-0.8174797638902463,-2.886473078971974);
\draw [line width=1pt,color=dtsfsf] (2.4965264536970224,-1.6635370948647248)-- (0,0);
\draw [line width=1pt,color=dtsfsf] (2.897030699891703,-0.7792388105613011)-- (0,0);
% \draw (-0.4718181818181806,4.435454545454502) node[anchor=north west] {$s=0$};

\begin{scriptsize}

\draw [fill=black] (0,0) circle (1pt);

\draw [fill=black] (0,3) circle (3pt);

\draw[color=black] (-0.035454545454544184,3.444545454545421) node[text width=1cm,align=center] {BPM 1\\$s=0$};

\draw [fill=rvwvcq] (-3,0) circle (3pt);

\draw[color=rvwvcq] (-3.653636363636362,0.4263636363636339) node {BPM 7};

\draw [fill=rvwvcq] (0,-3) circle (3pt);

\draw[color=rvwvcq] (0.582727272727274,-3.1918181818181477) node {BPM 5};

\draw [fill=rvwvcq] (3,0) circle (3pt);

\draw[color=rvwvcq] (3.528181818181819,0.59) node {BPM 3};

\draw [fill=rvwvcq] (-2.0853550021597353,2.1566860028681436) circle (3pt);

\draw[color=rvwvcq] (-2.835454545454544,2.4445454545454313) node {BPM 8};

\draw [fill=rvwvcq] (2.068318017144739,-2.1730302758945754) circle (3pt);

\draw[color=rvwvcq] (2.3463636363636375,-2.41) node {BPM 4};

\draw [fill=rvwvcq] (2.1197367239559193,2.1229027818326083) circle (3pt);

\draw[color=rvwvcq] (2.800909090909092,2.0445454545454353) node {BPM 2};

\draw [fill=rvwvcq] (-2.1213203435596424,-2.121320343559643) circle (3pt);

\draw[color=rvwvcq] (-2.9990909090909077,-1.8827272727272517) node {BPM 6};

\draw [fill=dtsfsf] (1.1369868191682282,2.7761954133377778) circle (3pt);

\draw[color=dtsfsf] (1.2372727272727286,3.353636363636331) node {BPM 2};

\draw [fill=dtsfsf] (2.897030699891703,-0.7792388105613011) circle (3pt);

\draw[color=dtsfsf] (3.3645454545454556,-1.0827272727272599) node {BPM 3};

\draw [fill=dtsfsf] (2.4965264536970224,-1.6635370948647248) circle (3pt);

\draw[color=dtsfsf] (3.146363636363638,-1.6645454545454357) node {BPM 4};

\draw [fill=dtsfsf] (-0.8174797638902463,-2.886473078971974) circle (3pt);

\draw[color=dtsfsf] (-1.108181818181817,-3.1645454545454216) node {BPM 5};

\draw [fill=dtsfsf] (-1.692354305339511,-2.477082337185993) circle (3pt);

\draw[color=dtsfsf] (-2.144545454545453,-2.8827272727272417) node {BPM 6};

\draw [fill=dtsfsf] (-2.921064536389272,-0.6836534021336591) circle (3pt);

\draw[color=dtsfsf] (-3.5809090909090893,-0.8281818181818079) node {BPM 7};

\draw [fill=dtsfsf] (-1.5800664498749424,2.550174506573932) circle (3pt);

\draw[color=dtsfsf] (-1.5809090909090897,2.93) node {BPM 8};

\end{scriptsize}\end{tikzpicture}
\caption{\label{Method:Hyp_Ring}A hypothetical ring with eight BPMs at asymmetrical longitudinal positions, which are shown with red circles. The mixed BPM method assumes that these BPMs, are actually at the locations with the blue markers where they divide the circumference of the ring in exactly eight equal parts. This introduces an error $\delta_k$ for each $k$ BPM. BPM $1$ is set as the reference point.}
\end{figure}
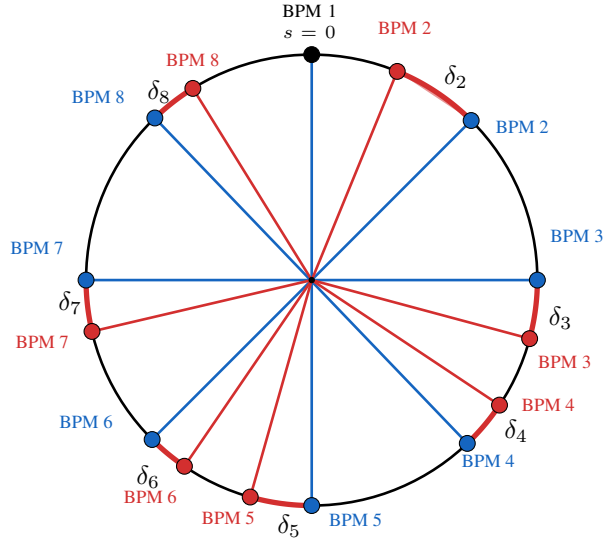

The transformation of the TbT data from $M$ BPMs and for $N$ turns, results in a vector of samples, where $N$ groups of $M$ samples each are created. The periodicity of each group is repeated for every turn. However, the mixed BPM method assumes that the BPMs, which are used for the reconstruction of the beam's trajectory in a BPM-by-BPM manner, are equidistant i.e. $M$ BPMs are situated at locations where they divide the circumference $C$ of the ring at exactly $\frac{C}{M}$ segments. In reality, the sampling intervals $\tau_k$, which are defined as the time interval that the beam needs to transport from BPM $k-1$ to BPM $k$, are not equal and they can be expressed as
\begin{equation}
\label{Method:tau_k_gen}
\tau_k=t_{k}-t_{k-1}\,\,,
\end{equation}
where $t_{k}$ and $t_{k-1}$ are the time instants that the beam passes from BPMs $k$ and $k-1$ respectively. The boundary condition $t_0$ can be chosen to be $t_0=0$. The periodicity conditions, due to the circular geometry of the ring, suggest that 
\begin{equation}
\label{Method:per_sample}
\tau_k=\tau_{k+M}\,\,.
\end{equation}

According to Eq.~\eqref{Method:tau_k_gen}, the time instances $t_k$ can be expressed for any $k$ as

\begin{equation}
\label{Method:tk_gen}
t_k=\sum_{n=1}^k \tau_n \,\,
%&=\frac{m}{M}T_o \,\,,
\end{equation}
and when the centroid of the beam passes from the BPM $k$ to the BPM $k+M$, it has circulated the ring once, i.e.
\begin{equation}
\label{Method:tau_per}
t_{k+M}-t_k=T_o \,\,,
\end{equation}
where $T_o$ is the revolution period of the beam. Combining Eq.~\eqref{Method:tau_per} and Eq.~\eqref{Method:tk_gen} yields
\begin{align}
\label{Method:ave_sample_per}
&\sum_{n=k+1}^{k+M}\tau_n=T_o\nonumber \\
&\frac{1}{M}\sum_{n=k+1}^{k+M}\tau_n=\frac{T_o}{M}\nonumber \\
&\langle\tau_k\rangle=\frac{T_o}{M}\,\,,
\end{align}
which means that, independently of the non-uniform positioning of the BPMs in the ring, the \emph{average} sampling period is bounded. Due to the periodicity condition Eq.~\eqref{Method:per_sample}, the expectation value $\langle\tau_k\rangle$ is constant as $k\to\infty$. The previous expression can be used to construct a relationship for the time instances $t_k$. By introducing an error $\delta_k$, which quantifies the deviation of the BPM $k$ from a longitudinally uniform position, the time instances for $k>0$ are simply

\begin{align}
\label{Method:tau_k_delta}
t_k&=k~\langle\tau_k\rangle+\delta_k\nonumber \\
&=k~\frac{T_o}{M}+\delta_k \,\,,
% &=k\frac{T_o}{M}~\biggl(1+\frac{M\delta_k}{kT_o}\biggr) \,\,,
\end{align}
where the errors $\delta_k$ are considered to be \emph{random independent variables}, given in units of seconds. In the case of single BPM analysis for $M=1$, $\delta_1=0$. The positions of the BPMs must not overlap, thus the error values are bounded in
\begin{equation}
-\frac{1}{2M}<\delta_k<\frac{1}{2M} \,\,.
\end{equation}
As a consequence, the expectation value of the $\delta_k$ errors as $k\to\infty$ is 
\begin{equation}
\langle\delta_k\rangle=0\,\,.
\end{equation}
Due to the periodicity of the ring, the errors are 1-turn periodic, i.e. $\delta_k=\delta_{k+M}$, which implies that the variance of $\delta_k$,  $\sigma_{\delta_k}^2$, is also bounded, for all $k$. The previous conditions highlight the fact the non-uniform sampling of the beam around the ring is \emph{stationary}, which allows the reconstruction of the trajectory of the beam simultaneously from all the BPMs. Therefore, the acquired signal, after the transformation of the TbT data with the mixed BPM method, can be considered as a band-limited signal with a non-uniform but \emph{recurrent} sampling scheme \cite{Yen_Sampling}.

The non-uniformity of the sampling process can be visualized in the fictitious ring of Fig.~\ref{Method:Hyp_Ring}, where $8$ BPMs (blue circles) are placed at locations where they divide the circumference of the hypothetical ring in $8$ equal parts. The injection point is assumed to be at the position of BPM 1 (black marker) and beam follows the clock-wise direction. As the centroid of the beam rotates around the ring during one turn, the $8$ fictitious and longitudinally symmetric BPMs (blue circles) sample the transverse coordinates with a constant sampling period of $\frac{T_o}{M}$.
The actual BPMs (red circles)  are situated in longitudinal positions that deviate from the symmetric positions by an off-set $\delta_k$ (red arc). This error is assumed to be positive for a real BPM which is downstream from the symmetric position and negative for a BPM which is upstream from the symmetric position. 

\subsection{\label{subsec:spectra_mixed}Frequency spectra}
\begin{figure}
  \includegraphics[width=\mysize\textwidth]{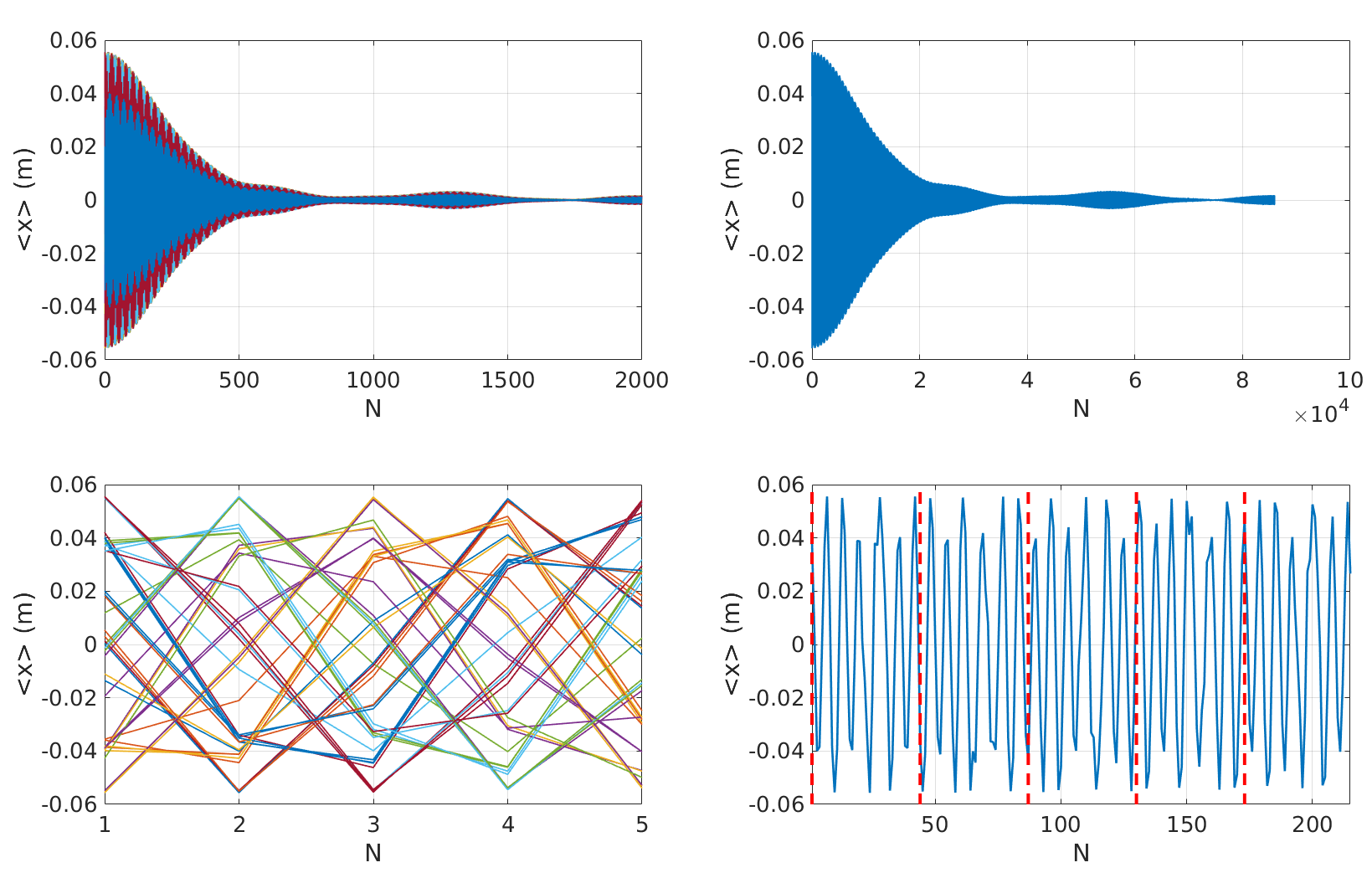}
  \caption{Synthesis of a one dimensional signal of 84000 samples (top right), from an initial two dimensional signal of N=2000 and M=42 (top left). The 5 first turns are shown (bottom left) with different color for each BPM. The transformation of the 5 turns results in 210 samples (bottom right), from the increase of the sampling rate. Each period is now populated by 42 samples of the original signal and the periods are shown between the red dashed lines.}
\label{Method:mixed_Example}
\end{figure}

The TbT data from multiple BPMs can be represented with an array. Let the array $A$, contain the spatial and temporal histories of the centroid of the beam:
\begin{align}
A =
\begin{bmatrix}
z_{11} & ... & z_{1M} \\
...     & ... & ... \\
z_{N1} & ... & z_{NM}
\end{bmatrix}\,\, ,
\label{matrix}
\end{align}
where z can be either $x$ or $y$ transverse planes, $N$ is the number of turns and $M$ the number of BPMs. Each column represents the signal from one BPM, with a sampling period of $\tau_s=1$ turn per sample. The traditional frequency analysis performs the tune measurements for each column of  the array in ~\eqref{matrix} and then provides an average tune estimation from the $M$ individual observations. Due to the correlations between the data in $A$, an increase of the sampling rate of the system is possible by vectorizing the array~\eqref{matrix} as:
\begin{equation}
\label{Method:signal}
\tilde{A}=[z_{11}z_{12}...z_{NM-1}z_{NM}]\,\, .
\end{equation}
An example of the transformation is shown in Fig.\ref{Method:mixed_Example}, where 2000 turns from 42 BPMs (top left) are transformed to a single vector of 84000 samples (top right). Indeed, the information of first 5 turns (bottom left) is transformed to a signal of 210 samples.

The vector $\tilde{A}$ contains $NM$ samples of the TbT data and the new sampling period becomes $\tilde{\tau_s}=\frac{1}{M}$, i.e. 1 turn per $M$ samples.

Although the sampling process of TbT BPM data is usually described in terms of  \emph{time}, the same procedure can be equally described in \emph{space}.
The mixed BPM signal of Eq.~\eqref{Method:signal} can be acquired by sampling the pseudo-harmonic oscillation 
\begin{equation}
\label{Method:eq1}
z(s)=\Re\bigg[\sqrt{\epsilon \beta(s)}e^{i(2\pi\Psi(s)+\phi_0)}\bigg]\,\,
\end{equation}
along the ring, where $s$ is the longitudinal variable, $\phi_0$ and $\epsilon$ are constants of the motion, $\beta(s)$ is the beta function and:
\begin{equation}
\label{Method:eq2}
 \Psi(s)=\int^s_0\frac{ds}{\beta(s)}\,\,,
\end{equation}
is the cumulative phase function with the constraint:

\begin{equation}
\label{Method:eq3}
%\Psi(0)&=0\nonumber \\
 \Psi(C)=Q_z\,\,,
\end{equation}
where $Q_z$ is the transverse betatron tune. The methodology consists of sampling Eq.~\eqref{Method:eq1} with $M$ BPMs, which are distributed at different longitudinal positions $\{s_1,s_2...,s_M\}$. Since the new signal consists of discrete samples, the continuous variable $s$ is dropped in favor of a discrete variable $m$. The generating function Eq.~\eqref{Method:eq1} becomes
\begin{equation}
\label{Method:eq_zm}
z(m)  = \Re\bigg[\sqrt{\epsilon \beta(m)}e^{i(2\pi\Psi(m)+\phi_0)}\bigg] 
\end{equation}
and the constraint $\Psi(M+1)=Q_z $.

The integral $\Psi(M+1)$ can be divided into $M$ equal parts and each part will advance the phase by a constant value of
\begin{equation}
\label{Method:eq5}
\Psi_0=\frac{Q_z}{M} \,\,,
\end{equation}
By introducing the aforementioned sampling errors $\delta_m$ in units of $2\pi$, for each $m$ sample, the cumulative phase $\Psi(m)$ is
\begin{align}
\label{Method:eq6}
\Psi(m)&=m\Psi_0+\delta_m\nonumber \\
&=m\frac{Q_z}{M}+\delta_m  \,\,.
%&=m\biggl(\frac{Q_z}{M}+\langle\delta\rangle_m\biggr)\,\,.
%\label{eq7}
\end{align}
With Eq.~\eqref{Method:eq6}, Eq.~\eqref{Method:eq_zm} is written as

\begin{equation}
\label{Method:eq7}
z(m) = \Re\bigg[\sqrt{\epsilon \beta(m)}e^{i (2\pi (m\frac{Qz}{M}+\delta_m)+\phi_0)}\bigg] \,\,.
\end{equation}
The periodicity of the beta function and of the error $\delta_m$, suggests that $\beta(m+M)=\beta(m)$ and $\delta_{m+M}=\delta_m$, i.e. the signal  Eq.~\eqref{Method:eq7} is modulated in amplitude and phase with a period of $M$.

In order to highlight the modulation of the signal, Eq.~\eqref{Method:eq7} is expressed as
\begin{align}
\label{Method:eq9}
z(m) & = \Re\bigg[\bigg(\sqrt{\epsilon \beta(m)}e^{i(2\pi \delta_m+\phi_0)}\bigg)e^{i2\pi m\frac{Qz}{M}}\bigg]\nonumber \\
& = \Re\bigg[H(m)e^{i2\pi m\frac{Qz}{M}}\bigg]\,\,.
\end{align}
The function $H(m)$ is $M$ periodic and it can be expanded in the Fourier series

\begin{equation}
\label{Method:eq10}
 H(m)=\sum^{+\infty}_{k=-\infty}C_k e^{\frac{i 2\pi m k}{M}}\,\,,
\end{equation}
with the weight functions for each harmonic defined as:

\begin{equation}
\label{Method:eq11}
C_k=\frac{1}{M}\int^{\frac{M}{2}}_{-\frac{M}{2}} H(m) e^{-i2\pi \frac{k m}{M}} dm
\end{equation}
Substitution of Eq~\eqref{Method:eq10} in Eq.~\eqref{Method:eq9} results in:

\begin{equation}
\label{Method:eq12}
z(m) = \Re\bigg[\sum^{+\infty}_{k=-\infty}C_k e^{i2\pi m(\frac{k+Qz}{M})}\bigg]
\end{equation}
Inspecting Eq.~\eqref{Method:eq12} the following observations can be made:

\begin{enumerate}[label=\roman*.]
\item Since the signal is modulated periodically in amplitude from the optics and in phase from the sampling error, infinite sidebands appear in the spectrum. The carrier frequency is the betatron tune which is found at the $k=0$ harmonic. The distance from the rest of the harmonics in the frequency spectra is $\frac{1}{M}$ which is the transformed sampling frequency. After transformation to the original frequency space by multiplying the frequencies with $M$,  this distance becomes 1. Therefore, the fractional betatron tune can always be recovered since the uncertainty between the harmonics is exactly 1 integer tune unit.

 \item The bandwidth of the signal in Eq.~\eqref{Method:eq10} is bounded by the Nyquist frequency. In the case of $M$ BPMs and according to Shannon's sampling theorem~\cite{Shannon}, the following relationship holds for the betatron tune:
 \begin{equation}
 \label{Method:Nyquist}
 Q_z\leq\frac{M}{2}\,\,,
 \end{equation}
 from where it can be deduced that, if the number of BPMs $M$ is at least twice the tune, then the integer part of the tune can be also recovered. Moreover, since the bandwidth of the signal has now become $M$ times larger, the usual discrepancy in the modulo of the fractional tune from single BPM analysis, is no longer present. This implies that tunes with fractional parts above 0.5 can be  recovered.
\end{enumerate}

\subsection{\label{Method:freq_res}Frequency resolution}
The frequency resolution i.e., the TbT error in the estimation of the betatron tunes, is correlated to the total observation time of a signal. For the mixed BPM method, the derived relationship of the time instances $t_k$, in the case of non equidistant $M$ BPMs with a sampling error $\delta_k$, is (see Section~\ref{Method_sub1})
\begin{equation}
\label{Method:t_k_res}
t_k=k\frac{T_o}{M}+\delta_k\,\,,
\end{equation}
with $T_o$ the revolution period. If the total number of samples is $m=MN$, the error $\delta_m=\delta_M$, since the last sample is sampled from the last BPM $M$. Moreover, Eq.~\eqref{Method:t_k_res} can be rewritten for $k=m$ as

\begin{equation}
\label{Method:t_k_res2}
t_m=m\frac{T_o}{M}g(m)\,\,,
\end{equation}
where $\tilde{\delta}_M$ is the sampling error for the BPM $M$, normalized to the revolution period and the function $g(m)$ is defined as
\begin{equation}
g(m)=\biggl(1+\frac{M\tilde{\delta}_M}{m}\biggr)\,\,.
\end{equation}
The expression in Eq.~\eqref{Method:t_k_res2} corresponds to the total observation time of the mixed BPM signal for $m=MN$ samples. This total time is now used for the cases of FFT and NAFF algorithms in order to estimate the error in the frequency analysis of each algorithm.
\subsubsection{FFT}

In the case of a simple FFT, the TbT error of the betatron frequency estimation for $m$ number of samples is

\begin{equation}
\label{Method:fft}
	\Delta\nu(m)=|\nu(m)-\nu_o| =\frac{1}{t_m},
\end{equation}
with $\nu(m)$ the time-dependent frequency estimation and $\nu_o$ the true frequency of the TbT data. The error $\epsilon(m)$ in the betatron tune $Q_o$ estimation within $m$ samples, is defined as 
\begin{equation}
\epsilon(m)=\Delta\nu(m)T_o=\frac{T_o}{t_m}
\end{equation}
and substitution of Eq.~\eqref{Method:t_k_res2} in the previous expression, with $m=MN$,  yields

\begin{equation}
\label{fft_conv}
	\epsilon(N) =\frac{1}{N+\tilde{\delta}_M}
%	&= \frac{1}{N}(1+M\langle\tilde{\delta}\rangle_{M}^T)^{-1}\,\,.
\end{equation}
Since the error $|\tilde{\delta}_M|<\frac{1}{2M}\ll N$, Eq.\eqref{fft_conv} can be expanded around $\tilde{\delta}_M\approx0$ to give
\begin{equation}
\label{fft_conv_2}
	\epsilon(N) =\frac{1}{N}-\frac{\tilde{\delta}_M}{N^2}+\mathcal{O}(\tilde{\delta}_M^2)
%	&= \frac{1}{N}(1+M\langle\tilde{\delta}\rangle_{M}^T)^{-1}\,\,.
\end{equation}
Clearly, the mixed BPM method when used with the FFT, does not result in any gain in convergence with respect to the single BPM analysis, $M=1$. Moreover, the sampling error has been introduced in the estimation of the tunes, although it converges faster to zero than $\frac{1}{N}$. It should be noted that the mixed BPM method, still allows the estimation of the integer part of the tune with a simple FFT, provided that the condition in Eq.~\eqref{Method:Nyquist} is fulfilled.

\subsubsection{NAFF}
In the case of NAFF algorithm, the signal under study is treated as \emph{quasiperiodic}~\cite{Moser}. The resolution of the tune measurements scales as $t_m^{-(2p+2)}$ for the total observation time $t_m$ in Eq.~\eqref{Method:t_k_res}.  For the limiting case of $\tilde{\delta}_{M}\to0$ and for $m=MN$ number of samples and by using the theory behind NAFF~\cite{LASKAR1993257}, the analytical relationship of the error in the estimation the tunes $\epsilon(N)$, is found to be (see Appendix~\ref{Appendix:NAFF}):
\begin{equation}
\label{Method:final_dq2}
\epsilon (N)=\frac{\midpoint{C_L}}{M^{2p+1}}~  \biggl(\frac{1}{N^{2p+2}}-(2p+2)\frac{\tilde{\delta}_M}{N^{2p+3}}\biggr)\,\,.
\end{equation}
where the factor $\midpoint{C_L}$ depends on the number of samples $m$, the error $\tilde{\delta}_M$ , the order of the Hann window $p$ and the betatron frequencies and amplitudes of the signal under study. In fact, in the case of rational frequencies and/or the presence of multiple harmonics in the signal, the factor $\midpoint{C_L}$ can diverge rapidly.
It is important to mention that the gain in resolution by a factor of $M^{2p+1}$ is followed by a blow-up of the factor $\midpoint{C_L}$, as it can be confirmed from the full derivation of Eq.~\eqref{Method:final_dq2} in the Appendix~\ref{Appendix:NAFF}. Moreover, due to the dependence of $\midpoint{C_L}$ on the order of Hann window, the blow-up is expected to increase for an increasing $p$. 

The transformation of the signal with the mixed BPM method, creates a signal with multiple harmonics around the main frequency line, as it has been shown in Section~\ref{subsec:spectra_mixed}. This behaviour can also interfere with the convergence of the betatron tunes estimations with a small number of turns $N$, due to the dependence of the $\midpoint{C_L}$ on the spectral quantity of the signal under analysis. As a result, collections of BPMs with the least modulation of the optics and the sampling period are expected to produce more precise betatron tune measurements for a very low number of turns $N$.
 
The contribution of the sampling error $\tilde{\delta}_M$ is negligible, since it is very small by construction and it rapidly approaches zero. Thus, for the mixed BPM method, the two previous effects that can make the factor $\midpoint{C_L}$ to diverge can potentially reduce the improvement in the estimation error of the betatron tunes.

\section{\label{sec:NumApplication}Numerical Simulations}

\begin{figure}[!htb]
	\centering
	\includegraphics[width=\mysize\textwidth]{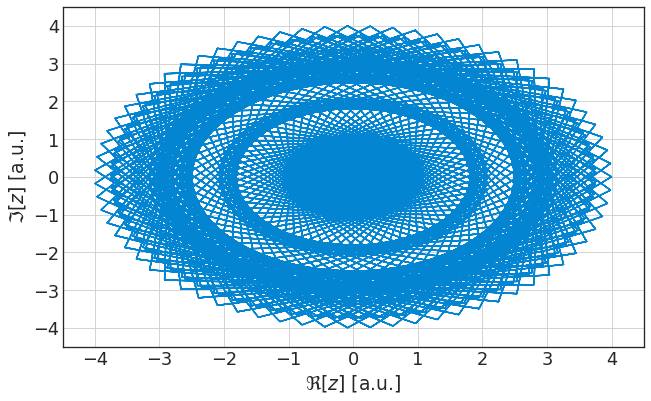}
	\caption{The projection of the signal used in the numerical simulations, on the complex plane. The oscillations consist of the superposition of four frequencies, with a frequency-to-frequency shift of $\Delta Q = 0.05$}
	\label{NumApp:model}
\end{figure}

Numerical simulations are performed with \emph{PyNAFF}~\cite{PyNAFF}, in order to qualitatively investigate the theoretical derivations of the mixed BPM method.  Since in a real machine, the TbT data from a BPM resemble pseudo-harmonic oscillations, the numerically simulated signal is chosen to be a superposition of four harmonic terms:

\begin{equation}
z(m)=\sum_{k=1}^4 e^{i 2 \pi Q_k (\frac{m}{M}+\delta_m)}\,\,,
\label{NumApp:signal}
\end{equation}
where $m$ is the index of each sample with the constraint $1\leq m \leq MN$, $M$ is the number of BPMs, $N$ the number of turns,  $Q_k$ is the tune of the harmonic $k$  in 2$\pi$ units and $\delta_m$ is the sampling error of sample $m$. The complex signal in Eq.~\eqref{NumApp:signal} contains information on the positions (real part) and the momenta (imaginary part) of the oscillations. In these numerical simulations, the real part of Eq.~\eqref{NumApp:signal}  is used, so as to simulate the types of signals that are acquired normally in an actual accelerator. The projection of the simulated TbT oscillations on the complex plane is presented in Fig.~\ref{NumApp:model}.
The values of the frequencies are chosen to be $Q_k=\frac{1}{2\pi}+(k-1) \Delta Q$ and the frequency separation  between the harmonics is $\Delta Q=0.05$.
%The ratio of the frequencies is always an irrational number, a characteristic of quasiperiodicity which is desired in this simulation.

The goal of the simulation is to use the mixed BPM method with NAFF, in order to measure the frequencies of the signal in Eq.~\eqref{NumApp:signal}. The uncertainty $\epsilon_k$ in the estimation of the $Q_k$ harmonic is defined as:
\begin{equation}
\label{NumApp:eps}
\epsilon_k(N,M) = \lvert Q_{k_o}-Q_k(N,M) \rvert\,\,,
\end{equation}
 where $Q_{k_o}$ are the actual tunes of the numerical signal Eq.~\eqref{NumApp:signal} and  $Q_k(N,M)$ is the frequency estimated over $N$  turns and $M$ BPMs. In this analysis, only the results of the $k=1$ harmonic will be presented, i.e. of the main frequency.
The $N$ variable is constrained to low values ($N\leq50$) in order to highlight the contribution of the parameter $M$ in Eq.~\eqref{Method:final_dq2}. Mixing the BPM data together, results in the increase of the sampling rate from one sample per turn to $M$ samples per turn. Therefore the frequencies $Q_k$ are transformed to $Q_k/M$  which will be referred to as the \emph{reduced frequencies}.

\begin{figure}[!htb]
	\centering
	\includegraphics[width=\mysize\textwidth]{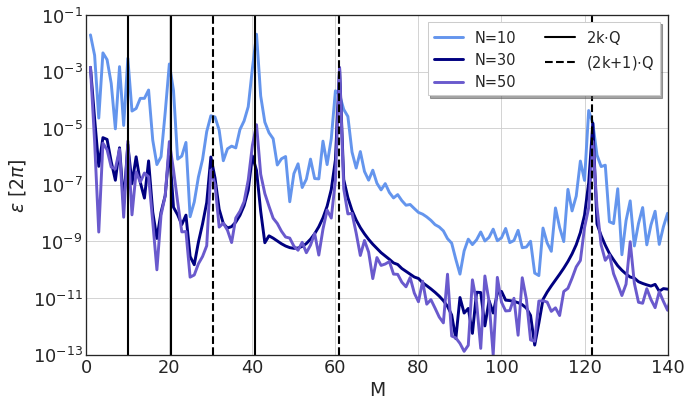}
	\caption{Appearance of resonances in the mixed BPM analysis of the numerical simulations, due to a rational $Q/M$. The measurement error $\epsilon$ is shown in logarithmic scale, for three cases of $N$, against the number of BPMs $M$. The even resonances are indicated with thick lines, while the odd resonances with dashed lines.}
	\label{NumApp:reson}
\end{figure}

For some $M$, the $Q_k/M$ ratio could yield a figure equal or almost equal to a rational number, which would result in the inability to reconstruct the original frequency with the convergence of Eq.~\eqref{Method:final_dq2}.  For example, for $M=20$, $Q_1/20\approx 1/2$, which is an even resonance.
 Indeed, this behavior is shown in Fig.~\ref{NumApp:reson}, where the mixed BPM method is applied and the error in the tune estimation $\epsilon_1$ is measured for an increasing number of BPMs $M$ and for three cases of $N$. Although the general trend shows a decreasing error with respect to $M$ for all cases of $N$, the convergence curves are contaminated due to the appearance of resonance peaks at specific numbers of BPMs $M$.
The even resonances are indicated with full lines, while the odd resonances with dashed lines. The comparison of the trends of the curves suggest a decrease of the error for an increasing  $N$ as expected, however the gain in convergence would be more evident in a "non-resonant" case.

In order to bypass these constraints and to demonstrate the convergence of the tune estimation with respect to the number of BPMs $M$, a \emph{varying} frequency is introduced, where for each case of $M$, the $Q_k$ values in Eq.~\eqref{NumApp:signal} are multiplied with $M$, in order to keep the $Q/M$ ratio constant.  In this way the generated TbT data do not lock on-to the aforementioned resonances and no systematic errors are introduced in the analysis.

\subsubsection{Tune convergence for increasing $M$}

\begin{figure}[!htb]
	\centering
	\includegraphics[width=\mysize\textwidth]{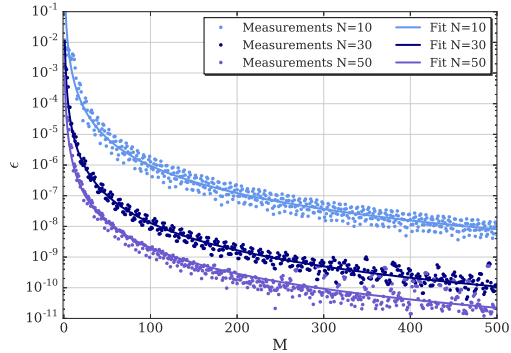}
	\caption{Convergenece of the mixed BPM method in the numerical simulations, for three cases of turns, with respect to the number of turns. After introducing the varying frequency scheme, the resonances disappear.}
	\label{NumApp:convergence}
\end{figure}

During the simulations, the sampling rate is kept constant and $\tilde{\delta}_m$=0, i.e. the $M$ BPMs are homogeneously distributed around the fictitious ring and the optics functions (betatron amplitude and phase advance) are equal at each BPM position. The order of the Hann window is set to be~$p=1$. The results of the mixed BPM measurements are shown in Fig.~\ref{NumApp:convergence}, where the error $\epsilon$ defined in Eq.~\ref{NumApp:eps} is plotted in logarithmic scale with respect to the number of BPMs. The same values of $N$ as in Fig.~\ref{NumApp:reson} are considered. Obviously the dependence of the error $\epsilon$ to the number of BPMs $M$ follows a power-law and a comparison with Fig.~\ref{NumApp:reson} confirms the absence of the resonance lines and the smooth convergence of the error $\epsilon$. For $N=10$ and $M=50$, the error is at the order of $10^{-5}$, while for $N=30$ and $N=50$, the error is around $10^{-7}$ and $10^{-8}$ respectively. These low errors are expected due to the use of a smooth quasiperiodic signal, Eq.~\ref{NumApp:signal}, the absence of additive noise in the signals and the constant sampling rate and optics. A fit of the convergence $\epsilon$ with a model of the form:
\begin{equation}
\label{NumApp:surf_1}
y(M)=c_1+c_2~log_{10}(M)\,\,,
\end{equation}
where$y(M)=log_{10}(\epsilon_1)$, $c_1$ is a constant term and $c_2$ the exponent of $M$ in the error $\epsilon$, yields $c_2=-3$, confirming the theoretical dependence of the convergence $\epsilon$ on $M$ for $p=1$, as it is shown in Eq~\eqref{Method:final_dq2} for $\tilde{\delta}_m$=0.

\begin{figure}[!htb]
  \centering
  \includegraphics[width=\mysize\textwidth]{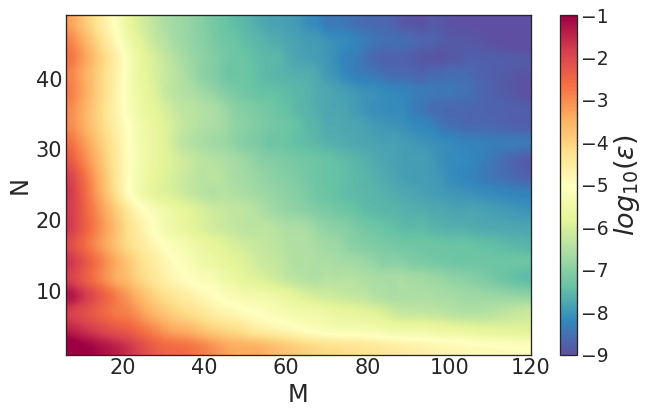}
  \caption{Colormap of the tune estimation error with the mixed BPM method for a numerical simulation with PyNAFF. Accurate values of the tune, at the order of $10^{-5}$ can be achieved either with a small number of turns ($N \leq 10$) and a larger number of BPMs ($M\geq 50$). Same conclusions can be made for $N \geq 40$ and $M \leq 30$.}
  \label{NumApp:colormap}
\end{figure}

% \newline
The role of the number of BPMs $M$ and turns $N$ in the error of the frequency estimation with NAFF is explored, by scanning over a range of ($M$,$N$) values. The resulting surface can be inspected in~Fig.\ref{NumApp:colormap}, where the color-bar represents the error $\epsilon$ in logarithmic scale and up to 50 turns are taken into account. The distribution of the convergence on the ($M$,$N$) surface appears to take hyperbolic shapes as a result of the power-law that it obeys. At first glance it is obvious that by increasing the number of BPMs, for a constant number of turns, the achieved error is gradually decreased. For example, below 10 turns the error reaches 10$^{-4}$ for $M=50$ and it decreases further at around 10$^{-7}$ for $M=100$. Same order of magnitude for the error can be achieved for $M=50$ and $N=40$ turns.

\subsubsection{Results for a varying window order}

The analytical relationship of the tune estimation error in Eq.~\eqref{Method:final_dq2}, suggests a strong dependence on the Hann window order $p$. In frequency analysis, a window is always used in order to reduce the impact of the finite sampling rate and finite duration of the signal, i.e. to compensate the leakage effect. For the case of a Hann window, the efficiency of the compensation depends strongly on the order of the window.  Although a higher order window leads to a faster damping of the error  $\epsilon$, studies have been performed~\cite{Kostoglou:2017iye} on the effect of different orders and the conclusions suggest that in the presence of noise in the BPM signal, e.g. from electronics, larger orders of the window lead to greater loss of precision.

\begin{figure}[htbp]
	\centering
	\includegraphics[width=\mysize\textwidth]{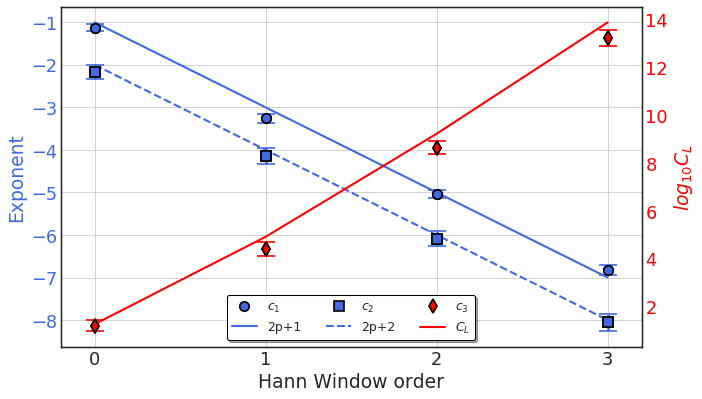}
	\caption{The coefficients estimated from the fit of the tune error $\epsilon_1$ to the surface of Eq.~\eqref{NumApp:surf}, along with their uncertainty. The $c_1$ coefficient is shown in blue circles, $c_2$ in blue squares and $c_3$ in red diamonds. The error-bars represent the 1$\sigma$ standard error of the fit. The theoretical predictions for the coefficients are shown with a blue dashed and thick line for $c_1$ and$c_2$ respectively. The theoretical estimation of $C_L$ is shown with a red line.}
	\label{NumApp:fitcoeff}
\end{figure}

For the numerical simulations presented in this section, frequency analysis is performed again for different values of ($M$,$N$) and for different values of the Hann window order $p$. The estimated convergence $\epsilon$ is fitted on the surface:

\begin{equation}
\label{NumApp:surf}
y(M,N)=c_1 log(M)+c_2log(N)+c_3\,\,,
\end{equation}
where $y(M,N)=log_{10}(\epsilon_1)$, $c_1=2p+1$,$c_2=2p+2$ and $c_3=log_{10}(C_L)$. The estimated values of the fit coefficients are presented in~Fig.\ref{NumApp:fitcoeff}, where each coefficient is shown for different orders of p, along with the theoretical expectations. Indeed, it is evident that the numerical results agree with the theoretical findings. The reconstruction of the TbT signal in Eq.~\eqref{NumApp:signal} with $M$ BPMs results in a reduction of the error by a factor of $M^{2p+1}$, as it is predicted by Eq.~\eqref{Method:final_dq2}. At the same time, the $C_L$ coefficient is increasing with respect to the window order $p$, as it is also expected from theory. 

\section{\label{sec:Sims}Tracking simulations}

The efficiency of the mixed BPM scheme is also tested on simulations with MADX-PTC~\cite{Schmidt:2002vp} and an optics model of the Proton Synchrotron (PS)~\cite{Benedikt:2000bs}. The parameters of the simulation can be found in Table \ref{table1}. No collective effects are considered in the simulations. The TbT data are recorded at 42 BPMs arranged around the PS lattice. The goal is to use the mixed BPM method for tune measurements with a very small number of turns $N\leq50$, make comparisons with the traditional single BPM measurement and explore the influence of the sampling rate error of $M$ BPMs on the resolution of the tune measurement.
% \begingroup
% \squeezetable
\begin{table}[htb!]
	\small
  \centering
  \caption{Parameters of the MADX-PTC tracking simulations with the PS Model.}
  \begin{tabular}{ll}
      \toprule
\textbf{Parameter} & \ \ \ \textbf{Value} \\
       \toprule
       Energy          &  \ \ \ 2.3 [$GeV$]        \\
       M           & \ \ \ 42 BPMs       \\
       Q$_x$, \ Q$_y$             & \ \ \ 6.24, \ 6.27 [2$\pi$]      \\
	   Q$'_x$, \ Q$'_y$             & \ \ \  -5.78, \ -7.66 [2$\pi$]       \\
	   Emittance $\epsilon_x$, \ $\epsilon_y$ & \ \ \  1.0, 0.8 \ [$\mu m rad$]  \\
	   Beam size $\sigma_x$, \ $\sigma_y$   & \ \ \  4.6, 3.0 [$mm$] \\
	   Dimensionality & \ \ \  4-D \\
	   Distribution & \ \  \ Gaussian \\
%	   No. of particles & \ \ \ 1000 \\

 % \bottomrule
 \toprule
  \end{tabular}
  \label{table1}
\end{table}
% \endgroup

\begin{figure}[htb!]
	\centering
	\includegraphics[width=\mysize\textwidth]{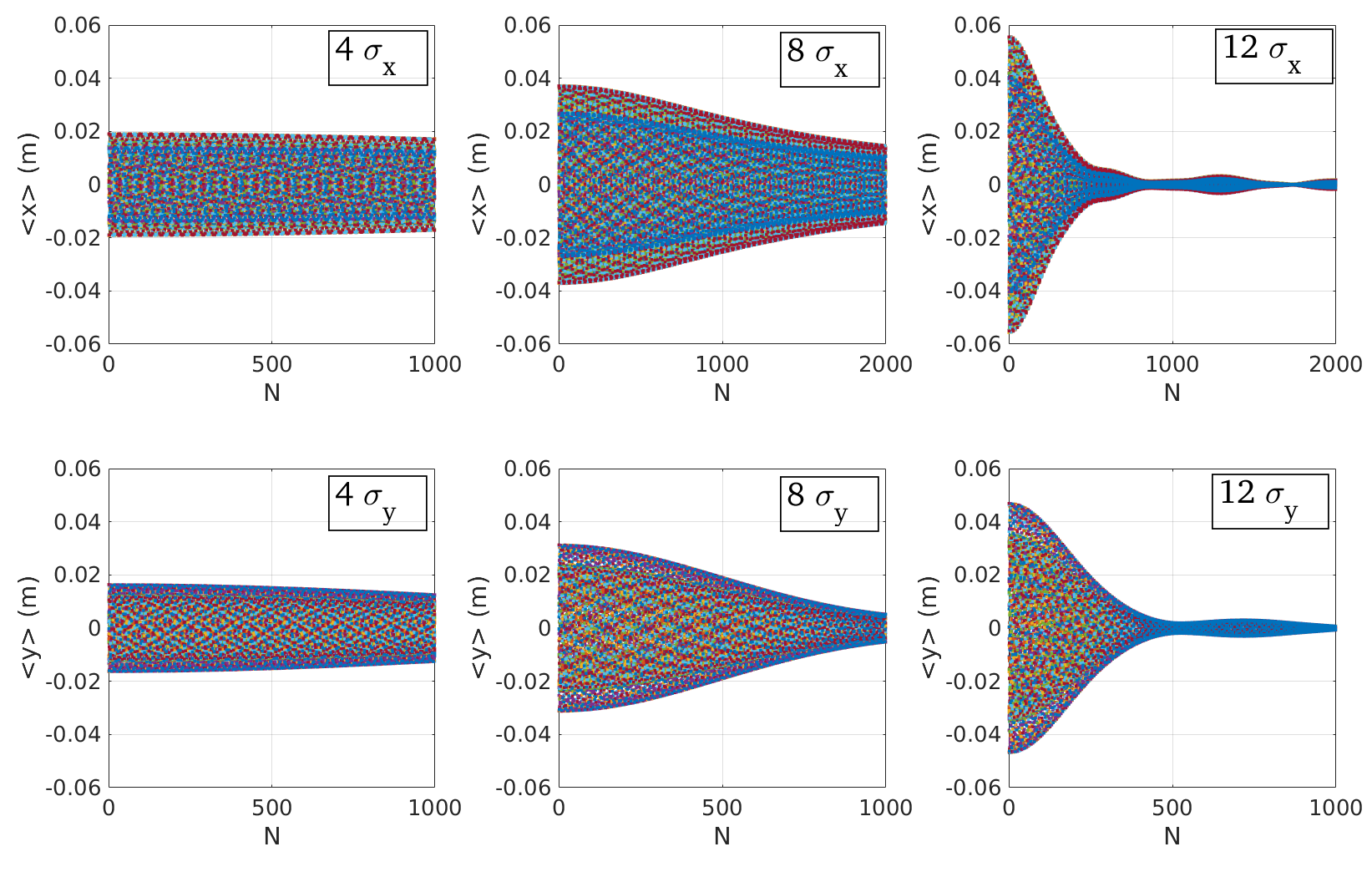}
	\caption{\label{sims:Deco} The turn by turn response of the centroid of the beam, for different kick strengths. Top row depicts the horizontal oscillation and the bottom row the vertical. Left column shows the TbT data for an initial amplitude of 4$\sigma$, middle column for 8$\sigma$ and right column for 12$\sigma$.}
\end{figure}

The beam is initially excited, horizontally and vertically, with deflections  that correspond to initial amplitudes of 2$\sigma$ to 12$\sigma$ with a step of 2$\sigma$. The centroid oscillations are measured from the average orbit of the particles. The evolution of the centroid of the beam is shown in Fig. \ref{sims:Deco} with respect to the number of turns $N$, for all the BPMs, and for the different initial conditions. The damping of the oscillations is faster for larger kicks, due to amplitude detuning coming from the non-linear magnetic elements. The beam exhibits maximum horizontal and vertical amplitude dependent tune-shifts of $(\Delta Q_x,\Delta Q_y)$=$(1.5\cdot10^{-3},2.5\cdot10^{-3})$ for excitations of 2$\sigma$ to 12$\sigma$.

\subsection{Optics and Symmetries}

\begin{figure}[htb!]
	\centering
	\includegraphics[width=\mysize\textwidth]{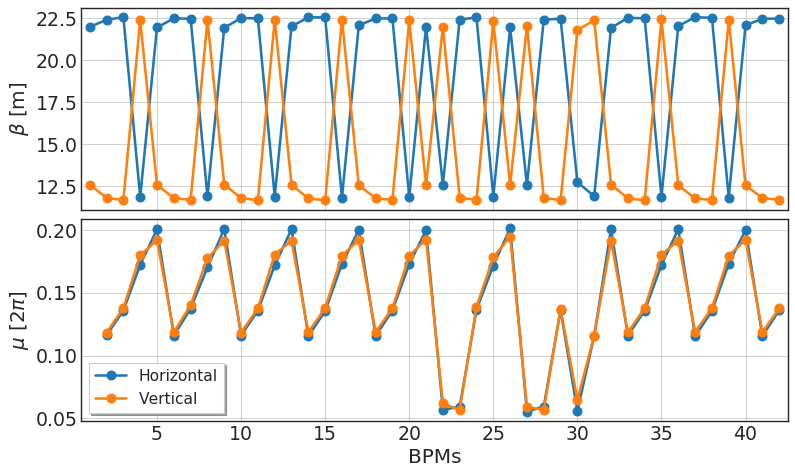}
	\caption{Beta function (top) and phase advance (bottom) with respect to the index of the 42 BPMs of the PS. The horizontal optics are shown in blue and the vertical in orange.}
	\label{sims:optics}
\end{figure}

The PS ring has a mean radius of $R=100$ m, and consists of $10$ super-periods, each made of $10$ combined function magnets. The optics used in these simulations corresponds to the so-called \emph{bare machine} optics, where the low energy quadrupoles which match the betatron tunes are not activated and there are no skew elements in the lattice. The beta functions and phase advances for both planes, at the location of the 42 BPMs are presented in Fig.\ref{sims:optics}, where the periodicity of the optics, for every 4 BPMs, is visible. This periodicity is however broken in the range of BPMs from 23 to 32. Since optics symmetry is a factor that can potentially interfere with the mixed BPMs method, two BPMs, which are located at positions that perturb the symmetry, are removed.

  Moreover, two interesting collections can be sampled from the new configuration of BPMs: a total of 30 BPMs that are almost symmetric with respect to their optics and a total of $10$ BPMs that are almost symmetric with respect to the optics and the longitudinal position.

The variation in the optics and in the longitudinal distance of all the above BPM configurations, is measured with the standard deviation and it can be seen in Table \ref{sims:table2}. From all the configurations, the one which exhibits symmetry in the optics and the longitudinal position is expected to produce the best precision and accuracy.
\begin{table}[htb!]
	\small
  \centering
  % \vspace{3pt}
    \begin{tabular}{ccccccc}
      \toprule
 Configuration & M  & $\sigma_{\beta_x}$ [m] & $\sigma_{\beta_y}$ [m] & $\sigma_{\mu_x} [2\pi]$ & $\sigma_{\mu_y} [2\pi]$ & $\sigma_s$ [m]   \\
       \toprule
       All     & 42 & 4.64 & 4.64 & 0.044 & 0.042 & 4.30 \\
       R.L.A.  & 40 & 4.53 & 4.49 & 0.030 & 0.030 & 4.27 \\
       O.S.    & 30 & 0.24 & 0.40 & 0.11 & 0.11 & 7.19  \\
       O.S. \& E.   & 10 & 0.052 & 0.013 & 0.00061 & 0.0012 & 0.45 \\

 %\bottomrule
 \toprule
  \end{tabular}
    \caption{\label{sims:table2}The variation of the optics (standard deviation) and the longitudinal distance between the used BPM configurations. R.L.A stands Reduced Lattice Asymmetry, O.S. for Optical Symmetry and E. for Equidistant.}
\end{table}

%\vspace{3pt}

\begin{figure}[htb!]
	\centering
	\includegraphics[width=\mysize\textwidth]{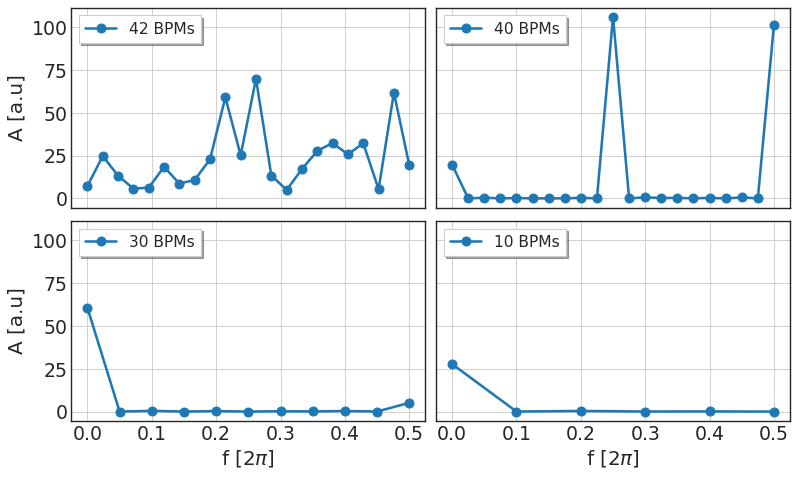}
	\caption{Fourier spectra of the horizontal beta functions for each BPM configuration signifying the level of modulation of the optics in mixed BPM method. The number of BPMs, associated to each configuration, is shown in the legend.}
	\label{sims:beta_mod}
\end{figure}

The Fourier spectra of the horizontal betatron function in one turn are shown in Fig. \ref{sims:beta_mod}, for all the aforementioned BPM collections. For the $42$ BPMs, two main peaks arise in the spectrum signifying the almost symmetric lattice of the PS. Reducing the number of BPMs to $40$, results in a cleaner spectrum with a very well defined oscillation frequency which corresponds to the 10$_{th}$ harmonic.

Sampling $30$ almost symmetric BPMs results in the spectra of the bottom left figure, where only low-frequency components are present. These slowly varying components appear due to the fact that the beta functions are almost, but not entirely, equal at the positions of these $30$ BPMs.  It should be noted that the closed orbit component has been removed from the data and the synchrotron motion is not included in the simulations. Finally, from the $10$ equidistant and symmetric to the optics BPMs, the spectrum also exhibits low frequency components but with half the amplitude of the $30$ BPMs case. The vertical beta function exhibits similar periodicities.

\begin{figure}[htb!]
	\centering
	\includegraphics[width=\mysize\textwidth]{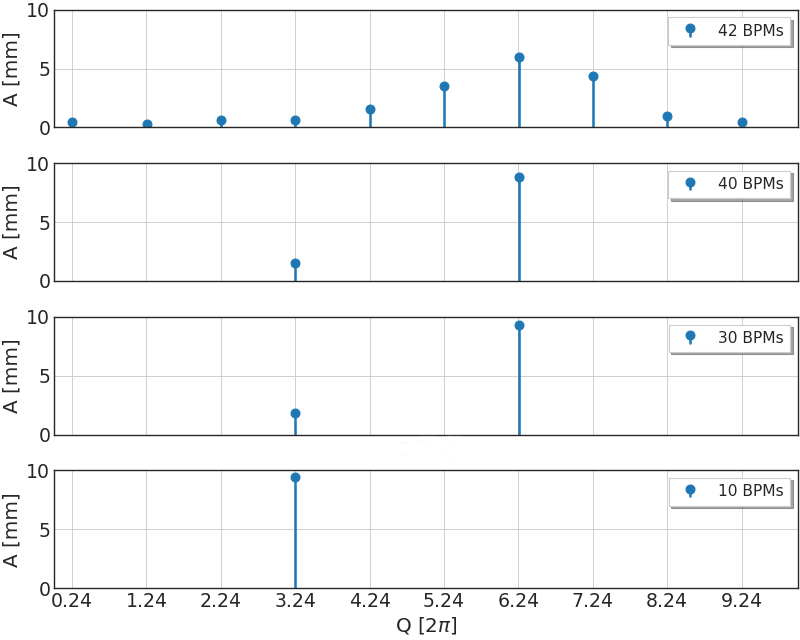}
	\caption{\label{sims:spectra}The Fourier Spectra of the mixed BPM signal for the different configurations of BPMs that are used in the analysis. The number of BPMs which corresponds to the different configurations is shown in the legend.}
\end{figure}

Due to the one-turn modulations of the optical and longitudinal asymmetries between the BPMs, sidebands are expected to appear in the Fourier Spectra of the mixed BPM signal. For the current chosen BPM configurations, the Fourier spectra for 24 turns are shown in Fig.\ref{sims:spectra}. On the top plot, the appearance of sidebands is shown around the main peak of betatron tune, with an exponentially decreasing amplitude.

By using the configuration of $40$ symmetric to the optics BPMs, only one sideband appears and the amplitude of the main peak increases due to the absence of any other spectral components, leading to a much cleaner signal. Furthermore, by using $30$ equidistant but not symmetric to the optics BPMs, the spectral quantity is almost identical. Moreover, by using $10$ symmetric and equidistant BPMs, the signal exhibits only one frequency. This component is the betatron frequency, however due to aliasing arising from the small number of BPMs, the integer part of the tune is found to be $3$ instead of $6$. Nevertheless, an important result is that the fractional part of the tune can always be determined, regardless of the symmetries between the BPMs.

\subsection{Tune Measurements}

\subsubsection{Precision}
\begin{figure}[htb!]
	\centering
	\includegraphics[width=0.5\textwidth]{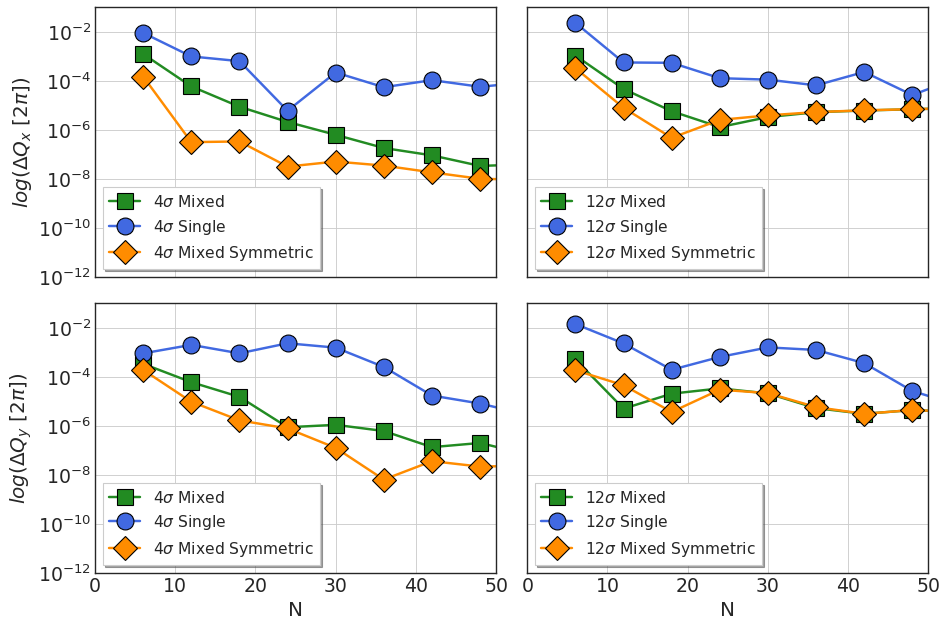}
	\caption{The precision of the tune measurements for  the horizontal (top row) and vertical (bottom row) planes. The results of the single BPM analysis is shown in blue circles, the mixed BPM scheme with 42 BPMs in green squares and the scheme with 10 BPMs, which are symmetric to the optics and to the longitudinal positions are shown in orange. The initial excitation of the beam corresponds to 4$\sigma$ (left column) and 12$\sigma$ (right column).}
	\label{sims:preci}
\end{figure}

A straightforward method to estimate precision is to measure the difference of TbT consecutive measurements. The results can be visualized in Fig. \ref{sims:preci}, where the TbT convergence of the measured horizontal (top row) and vertical (bottom row) tune values  is shown for the first 50 turns. Two cases of initial conditions are shown: the $4\sigma$ excitation (left column) and the $12\sigma$ excitation (right column). The different colors correspond to three BPM configurations. The case of a single BPM is shown in blue, the case of all the available 42 BPMs are shown in green and the case of the 10 equidistant and symmetric to the optics BPMs are shown in orange.

For the horizontal plane, the NAFF algorithm manages to estimate the tunes right from the first 6 turns for all BPM configurations and for both initial excitations. For the $4\sigma$ excitation, the single BPM precision is around 10$^{-2}$ at 6 turns, while the mixed BPM method exhibits a precision of 10$^ {-3}$ for 40 BPMs and even lower for the 10 equidistant and symmetric to the optics BPMs. At 50 turns, the precision is below 10$^{-7}$, whereas for the single BPM case it is at around 10$^{-4}$. Regarding the $12\sigma$ excitation, the mixed BPM method is found to be more precise than the single BPM analysis. However, for all configurations, the convergence of the measured betatron tunes is heavily modulated by the strong decoherence.

Same conclusions can be also drawn for the vertical tunes. Indeed, for the first 50 turns, mixing the BPM data together results in an improved precision. For the $4\sigma$ case, the equidistant and symmetric to the optics collection of 10 BPMs, exhibits the highest precision, while for the $12\sigma$ case, both mixed BPM collections present the same level of precision, almost $2$ orders of magnitude better than the single BPM case.

It should be noted that the improvement in precision of the transverse tunes estimation, is also found for the rest of the cases of initial excitations.

\subsubsection{Accuracy}

\begin{figure}[htb!]
	\centering
	\includegraphics[width=\mysize\textwidth]{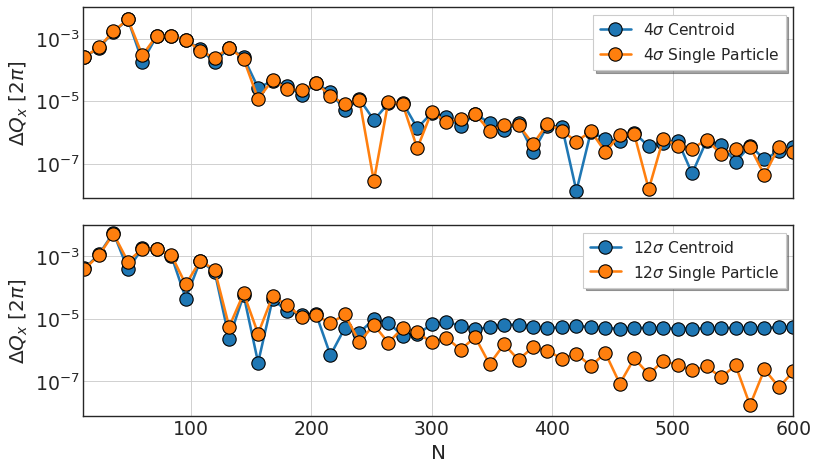}
	\caption{Comparison of the precision for the horizontal tune measurement of a single particle (orange) and the centroid of the beam (blue) with respect to the number of turns. Top figure corresponds to a 4$\sigma$ excitation while the bottom figure to a 12$\sigma$ excitation.}
	\label{sims:deco_effect}
\end{figure}

A common problem in estimating the accuracy of betatron tune measurements, is that the real value of the tune is not known a-priori. Concerning the present simulations, the reference value of the tune could be extracted from the transfer matrices of the particle tracking program, however this would introduce many systematic errors due to decoherence. An illustration of this effect is shown in Fig. \ref{sims:deco_effect}, where the precision of the tune measurements for the single particle and for the centroid of the particles distribution are plotted for the 4$\sigma$ excitation (top) and for the 12$\sigma$ excitation (bottom). The tunes are measured with single BPM analysis and the effect to be highlighted is that, while for the small excitation both convergences agree reasonably well, for the large excitation, a discrepancy is observed after 300 turns. After this point, decoherence due to amplitude detuning is dominant in the BPM signal.

 An intermediate solution to this problem could be to select a single particle which is located at the center of the transverse distributions and then measure the betatron tunes for a large number of turns. The single particle choice ensures that the only mechanism that would act on the tunes, is the tune-shift with amplitude with no dependence on the number of turns. In addition, a particle at the center of the distribution exhibits the least impact on its dynamics from the non-linear elements of the lattice. Under these considerations, the single particle tunes $Q_0$ are measured for all cases and compared with the measured centroid tunes $Q(N)$. Once more, comparisons between the single and the different mixed BPM configurations are made. 

\begin{figure}[htb!]
	\centering
	\includegraphics[width=0.5\textwidth]{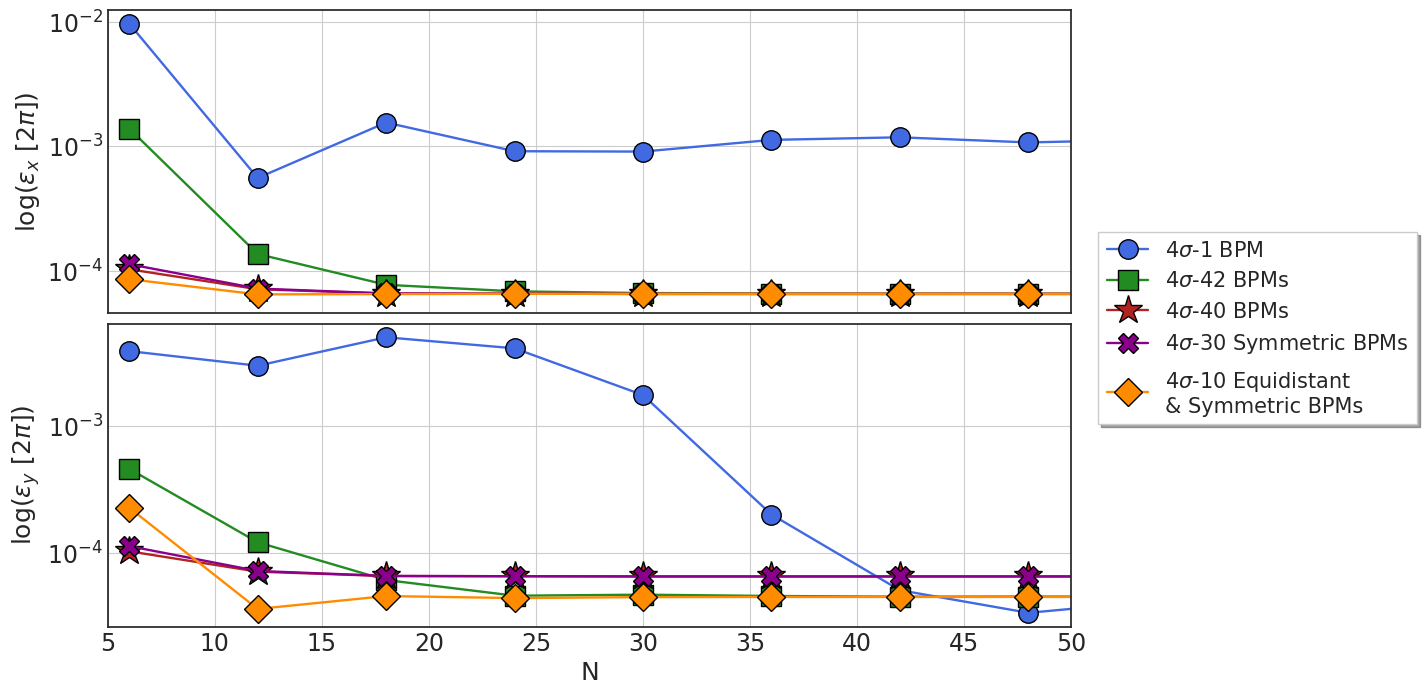}
	\caption{Comparison of the accuracy for the tune measurements with a single BPM (blue curve) and mixed BPM configurations of 42 BPMs (green curve), 40 BPMs (red), 30 BPMs (purple) and 10 BPMs (orange). The excitation corresponds to 4$\sigma$ for the horizontal (top) and vertical (bottom) tunes.}
	\label{sims:accu_4}
\end{figure}

The results for the 4$\sigma$ case are plotted in Fig.\ref{sims:accu_4}. In the top plot, the error in the horizontal tune estimation is shown with respect to the number of turns, where the improvement of accuracy for the mixed BPM cases is more than obvious. Even for the case of 42 BPMs, with no symmetry to the optics nor to the sampling frequency, the accuracy is at the order of $10^{-3}$ already at $12$ turns, one order of magnitude better than the single BPM case, while choosing symmetric BPMs, improves the accuracy for another order of magnitude. The effect of asymmetry in the 42 BPMs is reduced at the very first turns, as predicted from the relationship Eq.~\eqref{Method:final_dq2}. In this case, there is not an obvious gain between the 30 symmetric to the optics BPMs and the 10 symmetric to the longitudinal distance BPMs and to the optics BPM. In fact, there is only a marginal increase in accuracy at the very first turns for the latter configuration.

 In the bottom picture, the same estimation of the vertical tunes are shown, with similar improvement in the accuracy, when the mixed BPM scheme is used. The fact that the single BPM case in the vertical tune measurements exhibits a faster convergence of the error than the horizontal, can be attributed to the optics. Indeed, by inspecting the working point, the horizontal tune is very close to the 4$^{th}$ order resonance, which can be excited by strong sextupoles or octupoles. In any case, for both planes there is a substantial improvement in the tune estimation error for a very small number of turns, when the mixed BPM method is used.

\begin{figure}[htb!]
	\centering
	\includegraphics[width=0.5\textwidth]{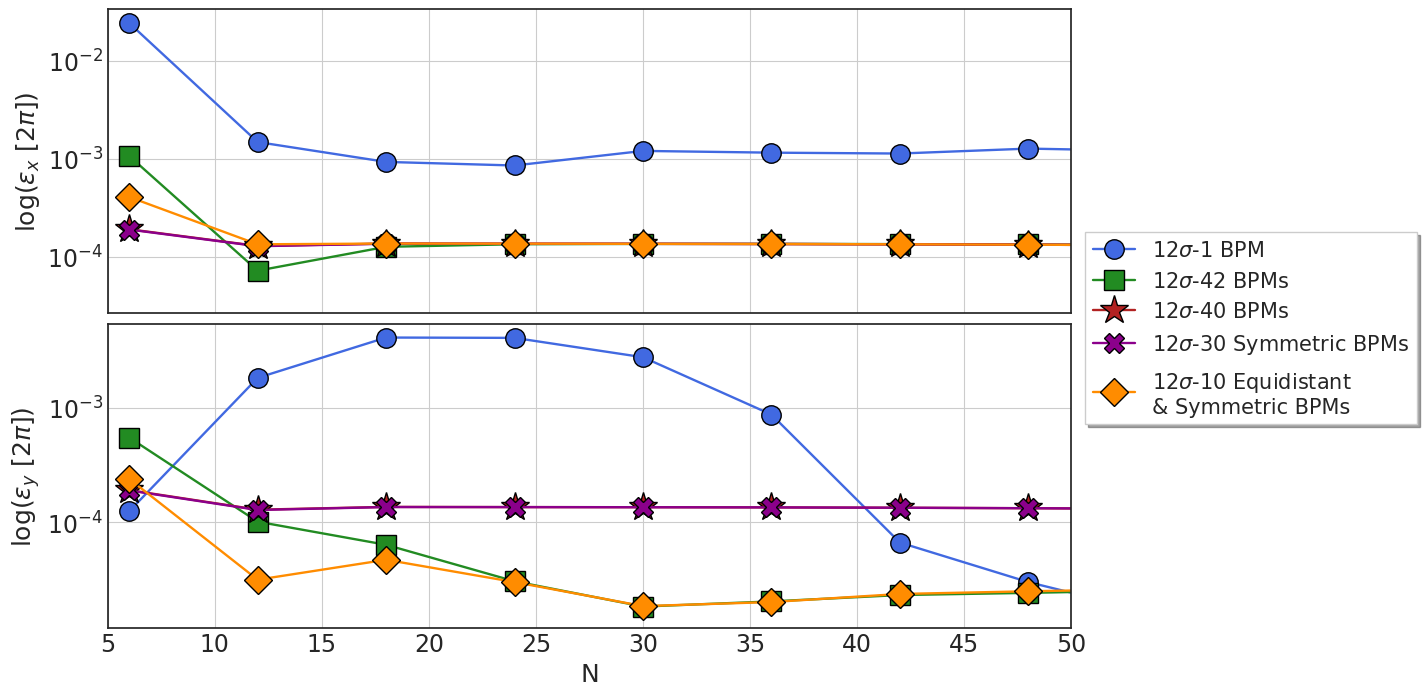}
	\caption{Comparison of the accuracy for the tune measurements with a single BPM (blue curve) and mixed BPM configurations of 42 BPMs (green curve), 40 BPMs (red), 30 BPMs (purple) and 10 BPMs (orange). The excitation corresponds to 12$\sigma$ for the horizontal (top) and vertical (bottom) tunes.}
	\label{sims:accu_12}
\end{figure}

Similar observations can be made in the case of a larger initial excitation. The tune error is shown for the case of 12$\sigma$ in Fig.\ref{sims:accu_12}, with the same configuration as in the previous case. Indeed, the mixed BPM method allows faster and more accurate determination of the betatron tunes in the very first turns than the single BPM method. For this case of large excitation, marginal differences can be observed in the tune estimation error, however they can be attributed to the fact that in this case, the impact of decoherence is stronger as compared to the 4$\sigma$ case.

\subsection{Equidistant BPMs}

The periodic modulation of the sampling period can result to loss of convergence for very small number of turns. In order to analyze this effect, a new collection of 10 BPMs which are not symmetric to the longitudinal position is sampled from the 42 BPMs. This collection is used for mixed BPM measurements and comparisons are made with the collection of 10 BPMs which are symmetric to the longitudinal position and to the optics. The collection of the longitudinally uniform BPMs is referred to as \emph{regular} and the non-uniform as \emph{irregular}. The normalized error $\delta_k$ in the sampling instance of the BPM $k$, can be estimated by using information of the longitudinal position of the BPMs, $s_k$. Indeed, the error is defined as

\begin{equation}
\tilde{\delta}_k=s_k-k\frac{C}{M}\,\,.
\end{equation}

\begin{figure}[htb!]
	\centering
	\includegraphics[width=0.5\textwidth]{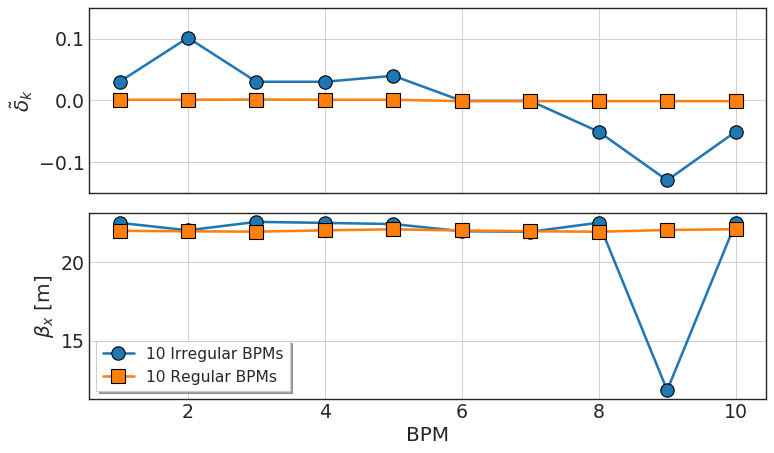}
	\caption{The normalized error $\delta_k$ (top) and the horizontal beta function (bottom) for 10 irregular BPMs (blue circles) and 10 regular BPMs (orange squares). The average va;ue of $\tilde{\delta}_k$ has been extracted.}
	\label{sims:accu_s_beta}
\end{figure}

The normalized error $\tilde{\delta}_k$ for the aforementioned BPM collections is shown in Fig.~\ref{sims:accu_s_beta} (top), together with the model horizontal beta function (bottom) for the irregular (orange line) and the regular (blue line) BPM collections. The oscillation of the error is negligible for the regular case, whereas for the irregular collection the oscillation of the error is obvious. In addition, the horizontal beta function of the regular case seems almost constant, while the irregular case exhibits larger deviations from a constant value. For the irregular case, both the root-mean-square (RMS) value of the error $\tilde{\delta}_k$ and of the horizontal beta function $\beta_x$, are almost 50 times larger than the respective RMS values of the regular case.

\begin{figure}[htb!]
	\centering
	\includegraphics[width=\mysize\textwidth]{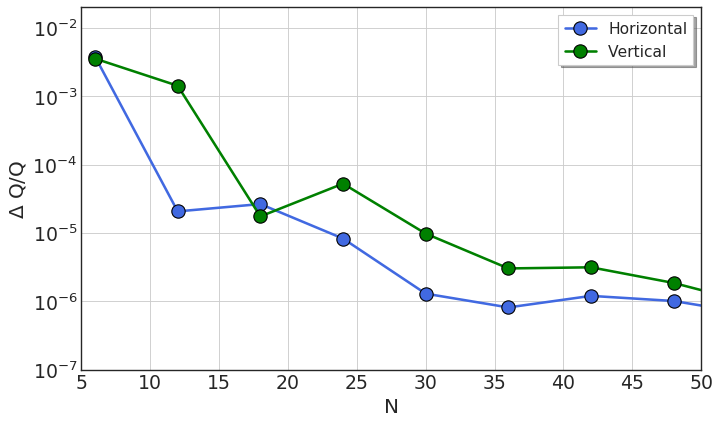}
	\caption{The normalized error between the tune measurements of 10 regular and 10 irregular BPMs, for the horizontal (blue curve)and the vertical (green curve) planes. The vertical axis is shown in logarithmic scale.}
	\label{sims:conv_ir_diff}
\end{figure}

The difference in the tune measurements of the two collections,relative to the tune measurements of the regular BPMs, are shown in Fig.\ref{sims:conv_ir_diff}, for the horizontal plane(blue) and the vertical plane (green) of the $4\sigma$ excitation. The trend of the curves exhibit a reduction of the relative difference and at $N$=30 turns, it is around 10$^{-6}$ for the horizontal and one order of magnitude larger for the vertical. The relative error converges at around $N$=50 turns. As a result, precise tune measurements are possible even for a collection of BPMs with large asymmetry in the longitudinal position.

\section{\label{sec:Applications}Applications on experimental TbT}

Over the past years, various measurements in proton and electron storage rings have been undertaken, where the mixed BPM method has been used with success for precise tune determination over a small number of turns \cite{Papaphilippou:2007zz,Zisopoulos:2015zkg, Zisopoulos:2017axr,Serluca:2017ltd}. Due to the unavoidable noise of the experimental TbT, any symmetries between the BPMs do not offer any substantial improvement in the convergence. This observation agrees with the detailed measurements that were performed at the ESRF by using the mixed BPM method~\cite{Papaphilippou:2007zz}. Some distinct cases of experimental measurements are presented in this section.

\subsection{PSB}
\begin{figure}[htb!]
	\centering
	\includegraphics[width=\mysize\textwidth]{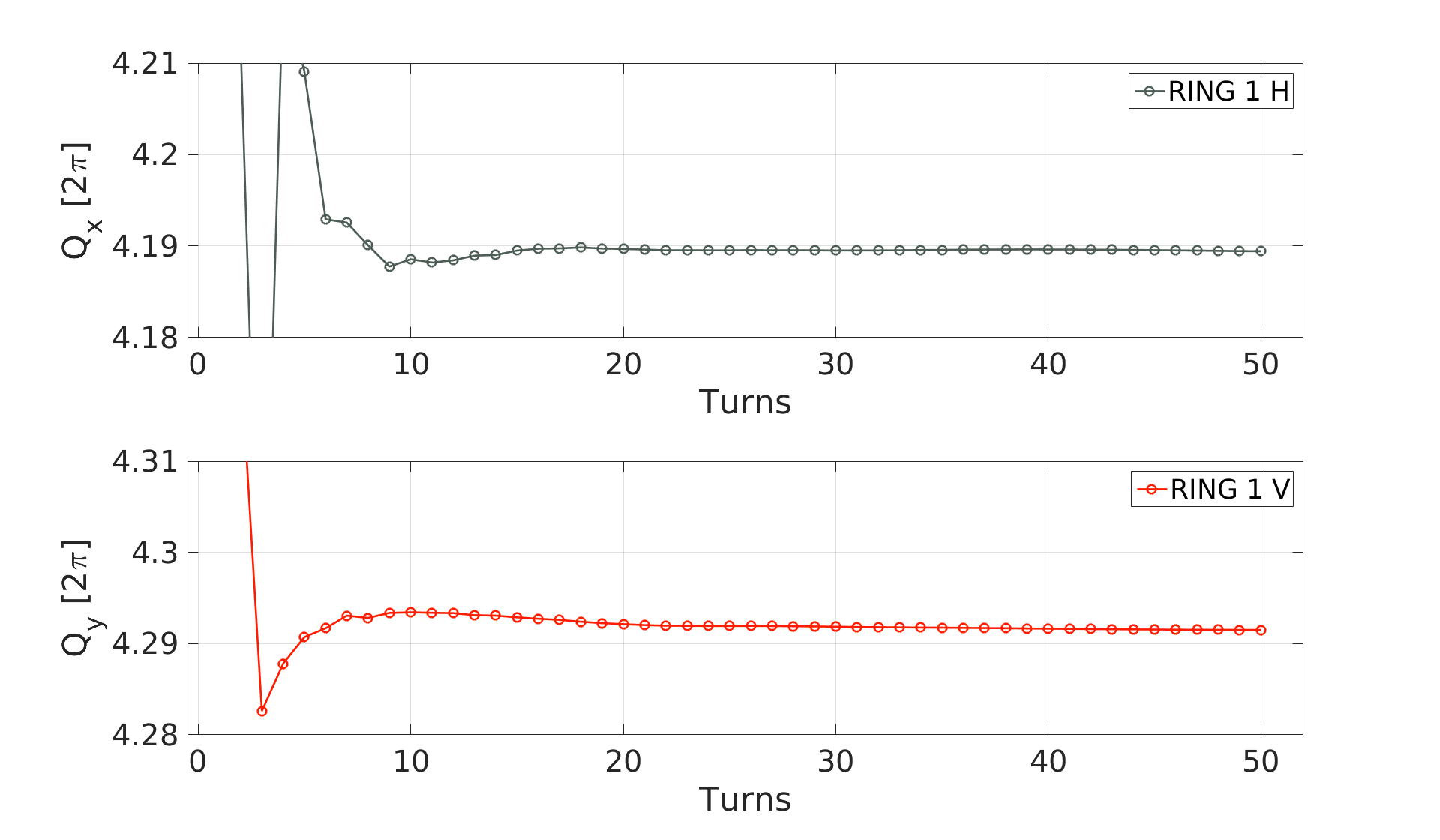}
	\caption{Tune measurements with respect to the number of turns N at the PSB Rign 1 by using horizontal (top) and vertical (bottom) data.}
	\label{exp:psb1}
\end{figure}

\begin{figure}[htb!]
	\centering
	\includegraphics[width=\mysize\textwidth]{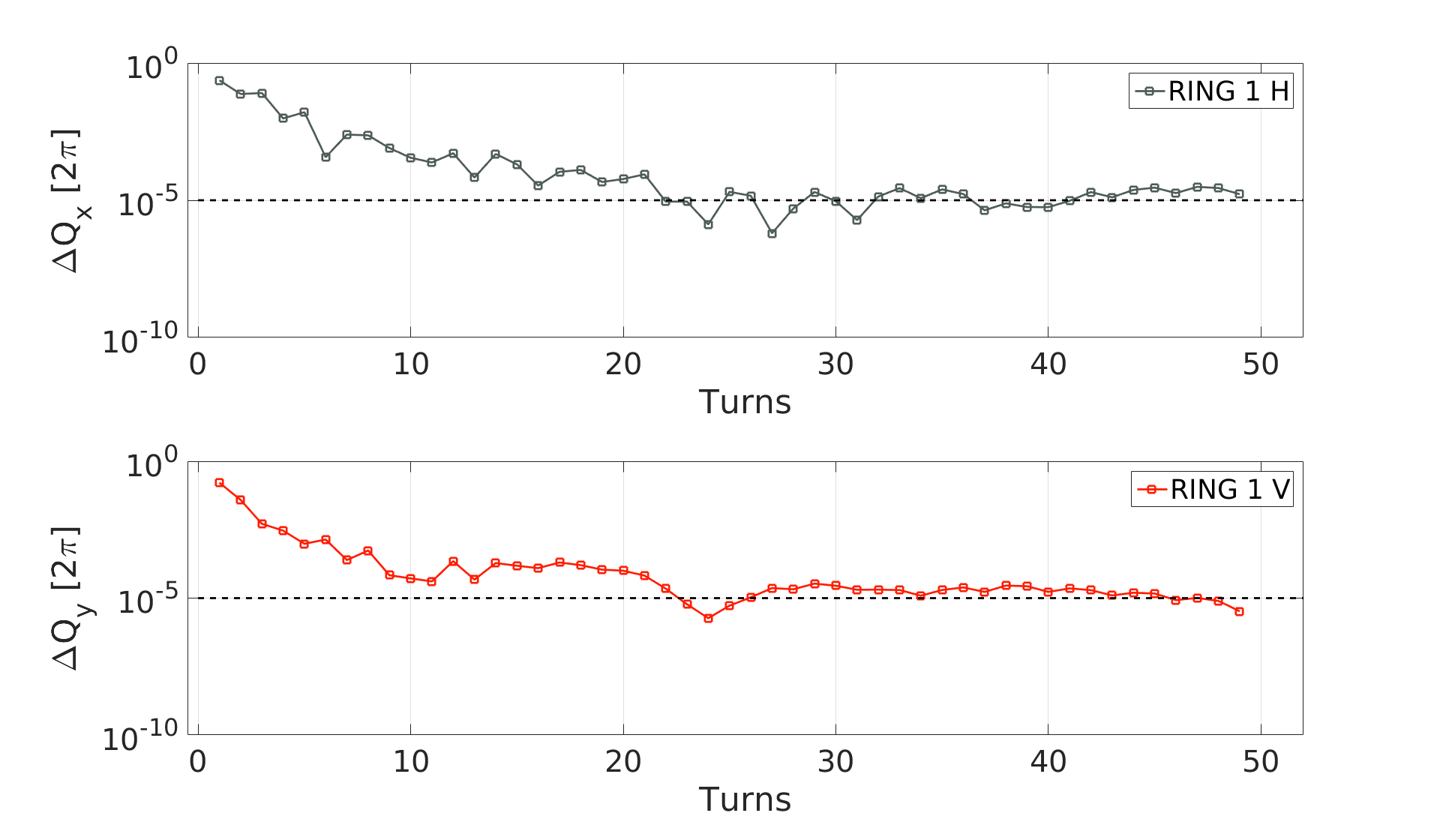}
	\caption{Convergence of the tune measurements at the PSB Ring 1 for the horizontal (top) and vertical (bottom) planes.}
	\label{exp:psb2}
\end{figure}

At the PSB, TbT data are analyzed from beams that are excited by a single kick, horizontally and vertically, at 160 MeV. Recently, the PSB has been undergoing many upgrades in view of the LIU Project~\cite{Hanke:2015gkg}. The ability for precise tune measurements is very important for such a low-energy machine, where collective effects and instabilities can drive the betatron tunes on resonances. The interesting feature of the PSB lattice is that it has a 16-fold symmetry and 16 BPMs, i.e. one BPM is located at each super-period and the optics exhibit no modulations. In addition, the BPMs are equidistant, with an almost equal longitudinal distance of about $\Delta_s$=10 m. As a result, precise measurements could be performed with the mixed BPM method, in a few turns. 
The results of horizontal and vertical tune measurements, from the aforementioned TbT data, are presented in Fig.\ref{exp:psb1} and Fig.\ref{exp:psb2}. The error-bars represent one standard deviation of the statistical uncertainty.  Remarkably, the betatron tunes can be extracted with only 10 turns and the convergence at 20 turns is about 10$^{-5}$. The experimental TbT had been also filtered from noise by using Singular Value Decomposition (SVD)~\cite{Wang:1999tt}, which helps as well in the fast and precise measurement of the tunes.

\subsection{KARA}

\begin{figure}[htb!]
	\centering
	\includegraphics[width=\mysize\textwidth]{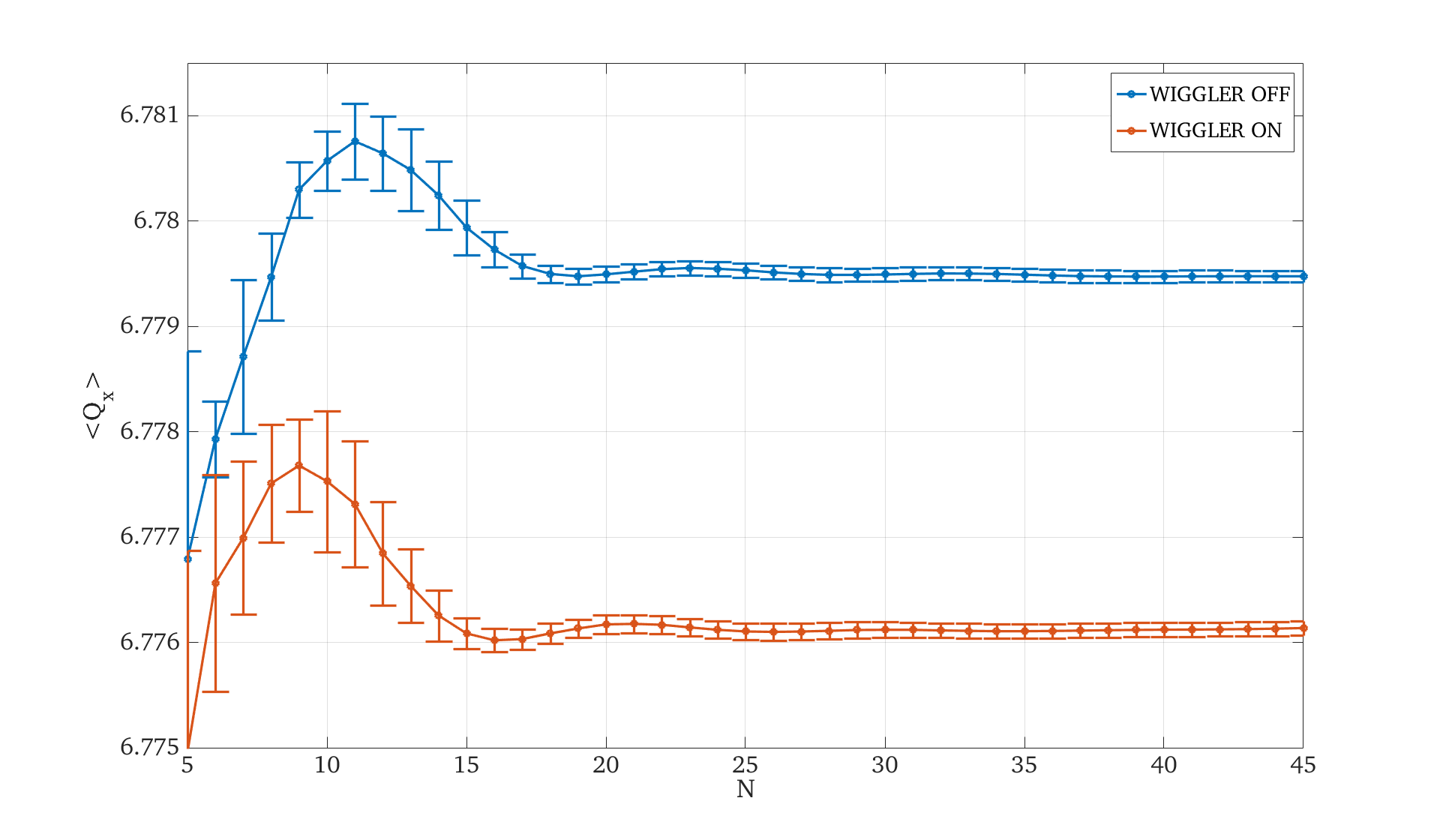}
	\caption{Betatron tune measurements at the ANKA light source for the horizontal (top) and vertical (bottom) planes. The case of operation under the influence of the CLIC Wiggler at 2.9 T is shown in red, while the zero field case is shown in blue. The error-bars are one statistical uncertainty of the measurements.}
	\label{exp:anka}
\end{figure}

The KARA ring (previously named ANKA) is a very flexible electron light source, operating at 2.4 GeV. The ring has a 4-fold symmetry and it is equipped with 39 BPMs. Recently, the prototype of the Superconducting Wiggler for the CLIC Damping Rings~\cite{Papaphilippou:2012zz,Bernhard:2016bzk} was installed at KARA in order to perform tests. In parallel, optics measurements have been performed with the wiggler at maximum field (2.9 T). The measurements of the horizontal tune with the mixed BPM method, are shown in Fig.\ref{exp:anka}, along with the statistical uncertainty. The method proved to be very efficient also in this case, since the horizontal tunes can be evaluated at around 20 turns, with a convergence of below 10$^{-3}$. This measurement is important in order to quantify any quadrupolar effects of the wiggler on the horizontal tune. Indeed, the results show that a slight horizontal tune-shift is present when the field of the wiggler changes from 0 T to 2.9 T. The tune-shift is at the order of about $\Delta Q$=4 10$^{-3}$.

\subsection{CERN PS}
The PS is one of the most indispensable parts of the Large Hadron Collider (LHC) Injector complex. In the PS ring, the final time structure of the bunches is shaped, a parameter which is very important for the efficient operation of the LHC. At injection, the PS receives bunches from the PSB at kinetic energy of 1.4 GeV and then it accelerates them up to 14-26 GeV. During injection, TbT data are gathered from 43 BPMs around the ring and the mixed BPM method is used in order to be benchmarked against the traditional tune measurement techniques. In order to investigate the injection process, which involves a very fast injection bump \cite{Serluca:2017ngg} and lasts for about 500 turns, a scanning window of 40 turns is applied on the data. In this way, any variations of the \emph{instantaneous} betatron tunes can be measured. The gating functionality of the BPM system at the PS, is used in order to perform the analysis bunch-by-bunch. A pre-analysis is also performed to characterize the mixed BPM method. With only 40 turns the convergence of the tunes is at the order of $10^{-4}$, which renders the method as very precise. It should be noted that similar levels of convergence cannot be reached with the traditional PS frequency analysis tools, for such a short time window.

\begin{figure}[htb!]
	\centering
	\includegraphics[width=\mysize\textwidth]{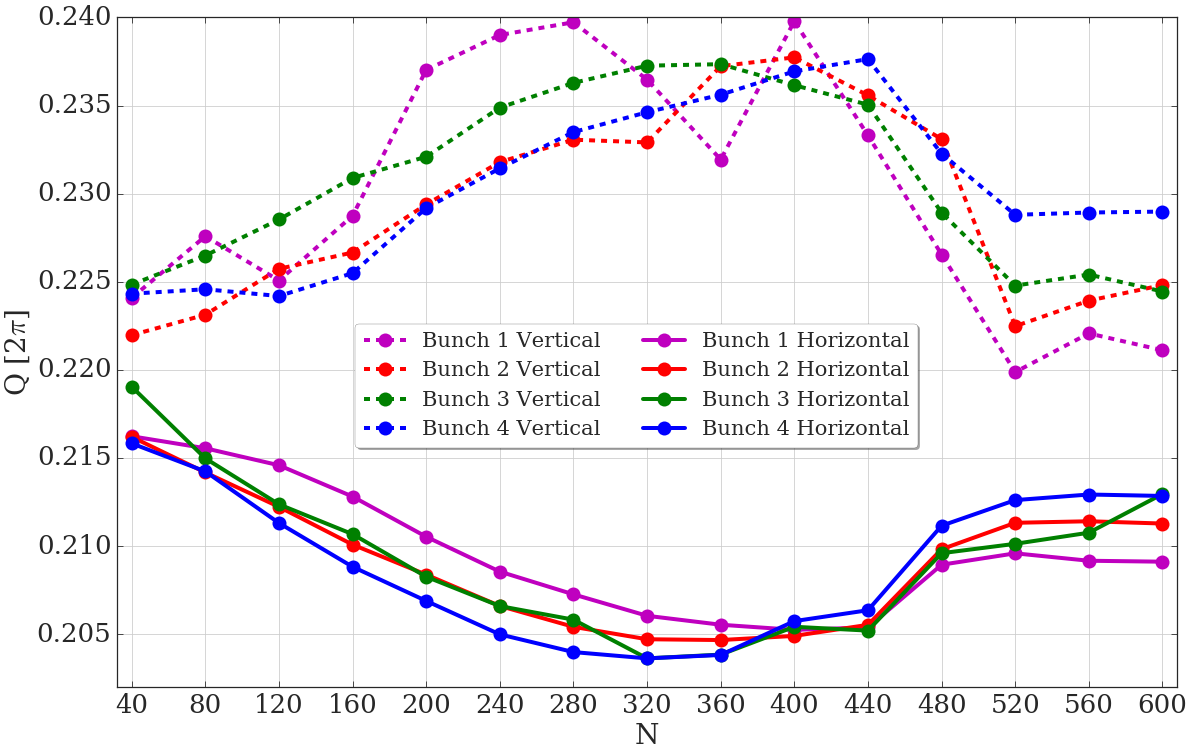}
	\caption{Instantaneous betatron tune measurements with the mixed BPM method, during the injection process at the PS. The estimation of the horizontal tunes shown in thick lines and of the vertical tunes in dashed lines. The analysis is performed for 4 bunches (bunch 1 in magenta, bunch 2 in red, bunch 3 in green and bunch 4 in blue) by using a sliding window of 40 turns.}
	\label{Appli:PS}
\end{figure}

The results of the scanning window analysis can be seen in Fig.~\ref{Appli:PS}, where the estimations of the horizontal (thick lines) and vertical (dashed lines) tunes for 4 bunches (bunch 1 in magenta, bunch 2 in red, bunch 3 in green and bunch 4 in blue) coming from the PSB, are shown. The observation of the periodic modulation of the tunes is evident. The depth of the modulation is such as to have a maximum tune-shift at the order of $10^{-2}$ for both planes. Due to the simultaneous presence of the effect in both planes and the large horizontal amplitude of the beam in the injection area, the effect is attributed to feed-down effects. More specifically, persistent sextupolar fields are created on the surface of the vacuum chamber of the injection bumpers and during injection, they modulate the quadrupolar component of the machine, resulting in the visible TbT tune-shift. This observation is significant for the LHC Injectors Upgrade (LIU) project \cite{Meddahi:2015tuo} because such large tune-shifts, can result to particle losses and reduction of the machine's efficiency as an injector. This effect could have not been evaluated without the possibility to estimate the tune and its evolution in a small number of turns, as provided by the mixed BPM method.

\section{\label{sec:Conclusions}Conclusions}
The efficiency of the mixed BPM scheme, when combined with NAFF, has been demonstrated for numerical,  tracking  and experimental data. For all cases, the method has been proven to be extremely efficient for precise tune measurements with a very small number of turns (below 50).

Due to amplitude and phase modulations, which are caused from the periodicity of the optics and of the sampling error at the locations of the BPMs, the estimation of the betatron tunes can be affected but results show that as the number of turns increases, this contribution is reduced. 

The mixed BPM method, which combines the TbT data from $M$ BPMs simultaneously, cannot be used with a simple FFT due to the linear dependence of the frequency resolution to the number of samples. When a Refined Fourier Analysis method is used, e.g. NAFF with a Hann window of order $p$, an improvement of $M^{-(2p+1)}$ is derived in the analytical expressions. Due to the periodic change of the sampling period at each BPM, an error is introduced in the tune estimations which is negligible and it is rapidly reduced.

Numerical simulations confirm the efficiency of the mixed BPM method, for precise measurements of the tune $Q$ within a small number of turns. The change of the sampling rate in the mixed BPM method results in the reduced tunes $Q$/$M$ and when this fraction is close a rational number, the error in the tune measurement is significantly increased. By implementing a varying frequency scheme, the expected dependence $M^{-(2p+1)}$ is recovered in these simulations.

The mixed BPM technique is also used for tracking data from the lattice of the PS machine. Comparisons with single BPM analysis validate the improvement in accuracy and precision, even for BPMs with asymmetries in the optics and/or the longitudinal position. Depending on magnitude of the modulation of the optics and the sampling rate of the $M$ BPMs, a decrease of the precision in the tune measurements is observed for a very small number of turns , however, this discrepancy is reduced quite fast, as the number of turns increases.

In the case of experimental data, maximization of the number of BPMs is often preferred, than using fewer BPMs with symmetries in the optics and/or to the longitudinal position. The reason for this preference is the existence of noise in the TbT data, which reduces significantly the efficiency of both the single and mixed BPMs methods. Nevertheless, the mixed BPM method has always proved to be very precise when experimental data are used. Useful applications of the mixed BPM method have been demonstrated on data from the PSB and PS proton rings, where the method has been proven capable to perform precise tune measurements in a very small number of turns. As a matter of fact, a large periodic tune-shift has been discovered recently at the PS, during the injection process, which could not have been observed with the single BPM analysis. The scheme of the mixed BPMs is also applied on data from the light source KARA, where betatron tune estimations help in the modeling of the beam's response to the CLIC Superconducting Wiggler.

The mixed BPM method has the potential to be a very useful tool in frequency analysis for accelerators as it offers a substantial improvement in the tune estimations. In addition,future studies could investigate improvements on the method, such as the \emph{interpolation} of non-uniformly sampled signals~\cite{Oppenheim:2011zz}, in order to minimize the contribution of the various periodic modulations at the position of the BPMs.

Finally, it should be noted that in cases of tracking simulations, the mixed BPM method can be particularly useful for many applications, e.g. for the construction of fast frequency maps or sliding windows in the TbT data to measure the instantaneous tunes. In cases of absence of noise from the data, schemes with BPMs that are equidistant or symmetric to the optics should be preferred, since they can provide even faster convergence, than using all the available BPMs.

% \begin{acknowledgments}
% We would like to acknowledge the PS and PSB OP teams and the KARA operators for being extremely helpful with the acquisition of the data. This work has been supported by the CLIC project and the Low Emittance Rings (LER) network of EUCARD2 and RULE of ARIES.
% \end{acknowledgments}

\section{Appendix A}
\subsection{Frequency resolution of the mixed BPM method with NAFF \label{Appendix:NAFF}}

The main success of NAFF is the ability to provide an approximation of a KAM~\cite{Dumas:1690640} quasi-periodic function $f(t)$, which can be a solution of a Hamiltonian system. The function $f(t)$ can be approximated as
\begin{equation}
\label{res:naff1}
f(t)\approx\sum^n_{k=1} \alpha_k e^{i \omega_k t}
\end{equation}
with $n$ the number of terms, $\alpha_k$ and $\omega_k$ the complex amplitude and real frequency of the $k_{th}$ harmonic.

Physically, the function $f(t)$ can be a trajectory of a non-linear quasi-periodic system, similar to the oscillation of the beam's centroid, along the lattice of an accelerator. The NAFF algorithm can recover the fundamental eigenfunctions of $f(t)$ in a very short time. For instance, the error in the estimation of the main frequency $\nu_1=\frac{\omega_1}{2\pi}$ of Eq.~\eqref{res:naff1} has been mathematically proven by Laskar to be~\cite{2003math......5364L}

\begin{equation}
\label{res:naff}
|\nu(T)-\nu_1|=\frac{C_L}{T^{2p+2}}+O(\frac{1}{T^{2p+2}})\,\,,
\end{equation}
for $T\to\infty$, where $T$ the total observation time, $\nu(T)$ is the time dependent frequency estimation of NAFF for the main harmonic, $p$ the order of the Hann window and the constant $C_L$ given as
\begin{equation}
\label{res:cl}
C_L=\frac{(-1)^{p+1}\pi^{2p}(p!)^2}{\phi''(0)}\sum^n_{k-(1,0...0)}\frac{\Re\{\alpha_k\}}{\Omega_k^{2p+1}}\cos(\Omega_k T)\,\,,
\end{equation}
where $\Omega_k=\langle \vec{k},\vec{\nu}\rangle- \nu_1$, with the basis vector $\vec{k}=\{k_1,k_2,...,k_n\}$, the frequency vector $\vec{\nu}=\{\nu_1,\nu_2,...,\nu_n\}$. Their inner product  $\langle \vec{k},\vec{\nu}\rangle=k_1\nu_1+k_2\nu_2+...+k_n\nu_n$ is the resonance space of the quasiperiodic solution,  $\Re\{\alpha_k\}$ the real part of the complex Fourier amplitude $\alpha_k$ of Eq.~\eqref{res:naff1} and the constant

\begin{equation}
	\phi''(0)= -\frac{2}{\pi^2}\biggl(\frac{\pi^2}{6}-\sum^p_{k=1}\frac{1}{k^2}\biggr)\,\,.
\end{equation}
Note that in the case of rational frequencies $\vec{\nu}$, the quasiperiodic orbits cannot be defined and $\Omega_k\to 0$, which makes the factor in~Eq.\eqref{res:cl} to diverge rapidly. In addition, the existence of multiple harmonics in the signal  Eq.~\eqref{res:naff1} can reduce the convergence to the actual frequencies as well.

In the practical case of frequency analysis of a quasiperiodic signal, where $m$ samples are gathered with a uniform sampling period $\tau_s$, Eq.~\eqref{res:naff} and \eqref{res:cl} are modified as:

\begin{equation}
\label{App:nu_m}
|\nu(m)-\nu|=\frac{C_L}{(m\tau_s)^{2p+2}}
\end{equation}
and the factor $C_L$ becomes:
\begin{equation}
C_L=\frac{(-1)^{p+1}\pi^{2p}(p!)^2}{\phi''(0)}\sum^n_{k-(1,0...0)}\frac{\Re\{\alpha_k\}}{(\tau_s\Omega_k)^{2p+1}}\cos(\Omega_k \tau_s)
\end{equation}
Next, the mixed BPM scheme is explored with the NAFF algorithm. Under this transformation, in the case of $M$ BPMs that record $N$ turns of the betatron oscillation of the beam, the total observation time is (see Sec.~\ref{Method_sub1})
\begin{equation}
\label{App:total_time}
T=m\frac{T_o}{M}+\delta_M\,\,,
\end{equation}
for $m$ number of samples. The term $T_o$ is the revolution period of the beam and $\delta_M$ is the deviation of the position of the last BPM from a fictitious position, which would be symmetric for all the $M$ BPMs. The previous expression can be used for estimating the error in the betatron tunes measurement with NAFF. 

The error in the betatron tunes estimation $\epsilon(m)$, with respect to the error in the frequencies $\Delta\nu(m)=|\nu(m)-\nu|$, for a total of $m$ samples is defined as

\begin{equation}
\label{App:eps}
\epsilon(m)=\Delta\nu~T_o\,\,.
\end{equation}
The total time in Eq.~\eqref{App:total_time} can be used to define a varying sampling period. Since the total observation time is $T=m\tau_s$ and from Eq.~\eqref{App:total_time} 
\begin{equation}
\label{App:total_time2}
T=m\frac{T_o}{M} g(m)\,\,,
\end{equation}
where $g(m)=1+\frac{M\tilde{\delta}_M}{m}$ and $\tilde{\delta}_M$ is the error of the last BPM normalized to the revolution period. Therefore, the sampling period $\tau_s$ can be expressed as
\begin{equation}
\tau_s=\frac{T_o}{M} g(m)\,\,.
\end{equation}
Substitution of the previous expression to Eq.~\eqref{App:nu_m} and combining the result with Eq.~\eqref{App:eps}, the error $\epsilon(m)$ in the betatron tune estimation for the mixed BPM method is written as

\begin{equation}
\label{Appendix:Las_med}
\epsilon (m)=\frac{M}{m^{2p+2}g(m)^{2p+2}}~\midpoint{C_L}(m),
\end{equation}
where the $\midpoint{C_L}(m)$ factor is

\begin{equation}
\label{Appendix:final_cl}
\midpoint{C_L}(m)=c_0\sum^n_{k-(1,0...0)} \frac{1}{\midpoint{\Omega_k}^{2p+1}} \Re(\alpha_k)\cos(mg(m)\midpoint{\Omega_k}) \,\,
\end{equation}
and the term $\midpoint{\Omega_k}$ is 

\begin{equation}
\midpoint{\Omega_k}=\frac{T_o~\Omega_k}{M}\,\,.
\end{equation}
In the case of $m=MN$ samples, the error $g(MN)$ is

\begin{equation}
g(MN)=1+\frac{\tilde{\delta}_M}{N}
\end{equation}
and Eq.~\eqref{Appendix:Las_med} becomes

\begin{equation}
\label{Appendix:Las_med_final}
\epsilon (MN)=\frac{\midpoint{C_L}(MN)}{M^{2p+1}N^{2p+2}}~  \biggl(1+\frac{\tilde{\delta}_M}{N}\biggr)^{-(2p+2)}\,\,.
\end{equation}
Since the error term $\tilde{\delta}_M \ll N$, the previous expression is expanded around $\tilde{\delta}_M\approx0$ which yields

\begin{equation}
\label{Appendix:final_dq2}
\epsilon (MN)=\frac{\midpoint{C_L}(MN)}{M^{2p+1}}~  \biggl(\frac{1}{N^{2p+2}}-(2p+2)\frac{\tilde{\delta}_M}{N^{2p+3}}\biggr)\,\,.
\end{equation}
The expression in Eq.~\eqref{Appendix:final_dq2} testifies that the convergence is improved by a factor of $M^{2p+1}$ with the mixed BPM method. The contribution of the small error $\tilde{\delta}_M$ converges to zero rapidly enough, so as to be negligible in the frequency analysis with NAFF and the mixed BPM method.

\bibliographystyle{unsrt}  
\bibliography{MixedBPM_arxiv.bib}% Produces the bibliography via BibTeX.

\end{document}